\documentclass[a4paper,11pt,headings=big,DIV=12]{scrartcl}
\pdfoutput=1
\usepackage{amsmath,amssymb,mathtools}

\usepackage[usenames, dvipsnames]{color}
\usepackage{tcolorbox} 
\definecolor{mygray}{gray}{0.2}
\addtokomafont{disposition}{\rmfamily\boldmath}
\usepackage{bbold}

\usepackage{jheppub}

\usepackage{bm,amssymb,slashed,graphicx,multirow,soul,mathtools,xspace,array}  
\usepackage{float}                   
\allowdisplaybreaks
\usepackage{ bbold }
\usepackage{subfigure}
 \usepackage{slashed}
 \usepackage{acronym}

\definecolor{violet}{rgb}{0.94, 0.2, 0.8}
\definecolor{lightblue}{rgb}{0.39, 0.58, 0.93} 
\definecolor{asparagus}{rgb}{0.53, 0.66, 0.42}

\newcommand{\FF}{form factor }
\newcommand{\FFs}{form factors }
\newcommand{\HA}{HA }
\newcommand{\HAs}{HAs }
\newcommand{\EQ}{Eq.~}
\newcommand{\EQs}{Eqs.~}

\acrodef{PDG}[PDG]{Particle Data Group}
\acrodef{OPE}[OPE]{Operator Product Expansion}
\acrodef{FCNC}[FCNC]{flavour-changing neutral current}
\acrodef{RHC}[RHC]{right-handed currents}
\acrodef{SM}[SM]{Standard Model}
\acrodef{NP}[NP]{New Physics}
\acrodef{MFV}[MFV]{Minimal Flavour Violation}
\acrodef{SD}[SD]{short-distance}
\acrodef{LD}[LD]{long-distance}
\acrodef{DA}[DA]{distribution amplitude}

\newcommand{\vev}[1]{\langle #1 \rangle}

\newcommand{\matel}[3]{\langle #1|#2|#3\rangle}
\newcommand{\al}{\alpha}
\newcommand{\be}{\beta}
\newcommand{\ga}{\gamma}
\newcommand{\de}{\delta}
\newcommand{\la}{\lambda}
\newcommand{\eps}{\epsilon}

\newcommand{\Cdot}{\!\cdot \!}
\newcommand{\mi}{\!-\!}
\newcommand{\pl}{\!+\!}

\newcommand{\TAB}{Tab.~}

\newcommand{\SEC}{Sec.~}

\newcommand{\APP}{App.~}
\newcommand{\APPs}{Apps.~}

\newcommand{\Rea}{\textrm{Re}}

\newcommand{\para}{\parallel}

\newcommand{\lone}{\ell_1}

\newcommand{\bltwo}{\bar{\ell}_2}

\definecolor{violet}{rgb}{0.94, 0.2, 0.8}
\definecolor{lightblue}{rgb}{0.39, 0.58, 0.93} 
\definecolor{lightgreen}{rgb}{0.1, 0.73, 0.33}

\DeclareOldFontCommand{\tt}{\normalfont\ttfamily}{\mathtt}

\newcommand*{\mathcolor}{}
\def\mathcolor#1#{\mathcoloraux{#1}}
\newcommand*{\mathcoloraux}[3]{%
  \protect\leavevmode
  \begingroup
    \color#1{#2}#3%
  \endgroup
}

\newcommand{\LAinb}{\Lambda_c}
\newcommand{\LAouts}{p}
\newcommand{\LAstarouts}{N}
\newcommand{\LAstaroutboth}{p(N)}

\newcommand{\LAin}{{\Lambda}}
\newcommand{\LAout}{\Lambda'} 
\newcommand{\LAstarout}{\Lambda^*}

\newcommand{\HAhad}[3]{H_{#1#2}^{#3}}
\newcommand{\HAlep}[3]{{\cal L}_{#1#2}^{#3}}

\newcommand{\fV}{f}
\newcommand{\fVbar}{\hat{f}}
\newcommand{\fT}{t}
\newcommand{\FV}{F}
\newcommand{\FT}{T}
\newcommand{\FVbar}{\hat{F}}

\newcommand{\momLab}{p} 
\newcommand{\momLap}{p_{\LAouts}} 
\newcommand{\momLapstars}{p_{\LAstarouts}} 
\newcommand{\momLapstarsh}{\hat{p}_{\LAstarouts}} 
\newcommand{\momLapboth}{p_{\LAstaroutboth }}
 
\newcommand{\momLabh}{\hat{p}} 
\newcommand{\spinLab}{\la_c} 
\newcommand{\spinLap}{\sigma} 
\newcommand{\spinq}{\la_q}

\newcommand{\massLab}{m_{\LAinb}}

\newcommand{\massLapsh}{\hat{m}_{\LAouts}}

\newcommand{\massLapstars}{{m}_{\LAstarouts}}
\newcommand{\massLapstarsh}{{\hat{m}}_{\LAstarouts}}
\newcommand{\massLapsboth}{{m}_{\LAouts(\LAstarouts)}}

\newcommand{\Q}[1]{s_{#1}}
\newcommand{\Qh}[1]{\hat{s}_{#1}}

\newcommand{\EPR}{E}
\newcommand{\Lst }{\Lambda^*}
\newcommand{\LN}{{\hat{L}}}
\newcommand{\KN}{{\hat{K}}}
\newcommand{\norma}{\frac{1}{\Omega}}
\newcommand{\normi}{{\Omega}}

\newcommand{\Mpl}{\Delta_+}
\newcommand{\Mmi}{\Delta_-}
\newcommand{\Mplh}{\hat{\Delta}_+}
\newcommand{\Mmih}{\hat{\Delta}_-}
\newcommand{\DelM}{\Delta^2}
\newcommand{\DelMh}{\hat{\Delta}^2}

\newcommand{\Mplmi}{\Delta_\pm}

\newcommand{\Mplmih}{\hat{\Delta}_\pm}
\newcommand{\Mmiplh}{\hat{\Delta}_\mp}

\newcommand{\qin}{c}
\newcommand{\qout}{u}

\newcommand{\qmaxsq}{q^2_{\textrm{max}}}

\newcommand{\aboost}{\eta}
\newcommand{\boost}{\be}
\newcommand{\nega}[1]{\bar{#1}}

\newcommand{\vv}{\kappa}
\newcommand{\gpol}{\omega}
\newcommand{\RS}{\Psi}

\begin{document}

\preprint{ \begin{flushright}  DO-TH 21/24 \end{flushright}
}

\title{\boldmath Endpoint Relations for  Baryons}

\author[1]{Gudrun Hiller,}
\author[2]{Roman Zwicky,}

\affiliation[1]{Institut f\"ur Physik, Technische Universit\"at Dortmund, D-44221 
Dortmund, Germany}
\affiliation[2]{Higgs Centre for Theoretical Physics, School of Physics and Astronomy, University of Edinburgh, Edinburgh EH9 3JZ, Scotland}

\emailAdd{gudrun.hiller@uni-dortmund.de}
\emailAdd{roman.zwicky@ed.ac.uk}

\abstract{Following our earlier work we establish kinematic 
endpoint relations for baryon decays using the Wigner-Eckart theorem 
and apply them to $\frac{1}{2} \to \frac{1}{2}$ and $\frac{1}{2} \to \frac{3}{2}$ baryon transitions. We provide  angular distributions at the kinematic endpoint which 
hold for the generic $d=6$  model-independent effective Hamiltonian 
and comment on  the behaviour in the vicinity of the endpoint.
Moreover, we  verify the endpoint relations, using an explicit form factor parametrisation,  and clarify constraints on helicity-based form factors to  evidence  endpoint relations.
Our results provide guidance for phenomenological parameterisations, 
consistency checks for theory computations and experiment.
Results are applicable to ongoing and future new physics searches at LHCb, BES III and Belle II with rare semileptonic-, dineutrino-and  charged-modes, which include
$\Lambda_b \to \Lambda^{(*)} \ell \ell, \Lambda_b \to \Lambda^{(*)} \nu \nu$, $\Omega_b \to \Omega \ell \ell$,  $\Lambda_c  \to p \ell \ell$,  $\Sigma \to p \ell \ell$
and  $\Lambda_b \to \Lambda_c^{(*)} \ell \nu$.}

\maketitle

\flushbottom

\setcounter{tocdepth}{3}
\setcounter{page}{1}
\pagestyle{plain}

\section{Introduction}
 
 \setcounter{page}{1}
 
The kinematic endpoint of a decay is characterised by the velocities,  of 
some of its decaying particles, approaching zero.  This results in the restoration of spherical symmetry 
 and leads to  enhanced symmetries in the  helicity amplitudes \cite{Z13}, 
 which crucially extends to effective field theories such as $\qin \to \qout \ell^+\ell^-$ or $b \to s \ell^+\ell^-$  \cite{HZ2013}.\footnote{Alternatively, it can be 
 viewed as the partial wave expansion applied to effective field theories.}  
 The ideas  behind this symmetry are completely general and apply at the non-perturbative level and any decay types including  the baryonic ones which are the focus of 
this presentation. 
At the level of the differential decay rate the relations result in exact predictions at the kinematic endpoint \cite{HZ2013}.  We refer to these predictions  as ``endpoint relations". 
On a pragmatic level they provide guiding for fit-parameterisation,   non-trivial consistency checks  for theoretical computations and  experiment itself.

The specific presentation of this paper applies to the baryonic decays\footnote{The baryons subscripts denote the $J^P$ quantum numbers and  the adaption to the case where
the initial state baryon is $J^P = \frac{1}{2}^-$ is straightforward as one can simply perform a parity transformation.}
\begin{alignat}{2}
\label{eq:cases}
& \LAin_{\frac{1}{2}^+} \to \LAout_{\frac{1}{2}^{\pm}} \, \lone \bltwo  \;, \quad   & & \textrm{e.g.  } 
\Lambda_c \to p \ell^+\ell^-   \;, \nonumber \\[0.1cm]
& \LAin_{\frac{1}{2}^+} \to \LAstarout_{ \frac{3}{2}^{\pm}}  \,  \lone \bltwo  & & \textrm{e.g.  } 
\Omega^0_c \to \Omega^- \ell^+ \nu   \;,
\end{alignat}
described by a $d=6$ effective Hamiltonian.\footnote{There are two loopholes. Firstly, 
QED-corrections break 
the factorisation into a hadronic and leptonic helicity amplitude (cf. section 5 in \cite{GHZ2016} for a discussion thereof and brief comments in \SEC\ref{sec:vic}). Second, new light-physics beyond the Standard Model  for which the effective theory approach 
becomes suboptimal. Whereas the first are presumably not too large the latter 
are an exciting possibility for which  endpoint relations are yet another tool to search for them.}
This includes decays of the semileptonic charged and 
flavour changing neutral current (FCNC)
type into charged leptons and neutrinos.
Such modes have been known for a long time as sensitive probes of  flavour physics  in  and beyond the Standard Model.
A vast number of theoretical and experimental studies for $K,D,B$-meson decays exists \cite{Buchalla:2008jp,HFLAV:2019otj}.

Rare decays of beauty, charm and strange baryons, such as 
$\Lambda_b \to \Lambda \ell \ell$ decays offer new  ways to test the SM and look for new physics,  complementing searches with mesons.
Due to  different theoretical and experimental systematics, caused by the interplay of strong \& weak dynamics and production \& reconstruction efficiencies respectively.
In addition, baryon modes provide opportunities on their own, including polarisation studies based on polarized baryon production
or angular distributions of the decay products of the final baryon.

Recently, observables induced by FCNC $b \to s \ell^+\ell^-$ transitions have been reported that hint at new physics and
specifically new phenomenon, violating lepton flavour universality \cite{Hiller:2014qzg,Bifani:2018zmi}.
This includes studies with $\Lambda_b$-decays \cite{LHCb:2019efc}, consistent with the flavour anomalies evidenced in $B$-decays \cite{LHCb:2021trn}, although with 
currently larger uncertainties.
Cross-checking findings from $B \to K^{(*)} \mu^+ \mu^-$ decays with  other modes induced by the
same FCNC transition provides a natural  and important next step to understand the origin of the flavour anomalies.
It is  therefore timely  to put baryonic modes at par with the mesonic ones.
For baryonic modes there are a number of form factor computations, 
e.g. lattice \cite{Detmold:2016pkz,Meinel:2016dqj,Zhang:2021oja,Meinel:2021mdj,Bahtiyar:2021voz} 
and sum rules \cite{Azizi:2011mw,Wang:2015ndk},
but no long-distance computations in QCD-factorisation or light-cone sum rules.

Rare baryonic decays are suitable for experimental study at high luminosity flavour facilities, such as LHCb \cite{Cerri:2018ypt}, Belle II \cite{Belle-II:2018jsg}, BES III \cite{BESIII:2020nme}, and possible future machines \cite{FCC:2018byv}.
Efforts exist for Hyperons \cite{Alves:2018npj} and in charm \cite{BaBar:2011ouc,LHCb:2017yqf,Meinel:2017ggx,Golz:2021imq}.
In beauty, the situation is more advanced, and first studies with angular distributions for  $b$-baryon decays are available \cite{LHCb:2018jna},
as well as  prospects for decays to spin $3/2$ baryons, such as $\Lambda_b \to \Lambda(1520) \ell \ell$  \cite{Amhis:2020phx}.
The $ \Lambda(1520) $ has received attention as it is the most pronounced resonance decaying strongly to $p K$ \cite{Legger:2006cq}.

The paper is organised as follows. In \SEC\ref{sec:mainidea} we review the ideas behind
the endpoint relations for helicity amplitudes from various viewpoints.
In \SEC\ref{sec:baryons} the concepts are applied to the  decays  in \eqref{eq:cases} 
at the level of the form factors. Moreover constraints on helicity-based form factors 
are discussed in \SEC\ref{sec:constraints} which are needed to evidence form factors relation  in this parametrisation.  In \SEC\ref{sec:AD} we apply the endpoint relation 
to angular distribution and derive exact relations valid at the endpoint.
The paper ends with summary and conclusions in \SEC\ref{sec:conclusions}.
\APPs \ref{app:conventions} and  \ref{app:t} contain conventions 
and a form factor derivation respectively.

\section{The Main Ideas behind Endpoint Relations and Symmetries}
\label{sec:mainidea}

We would like to present three ideas to understand the endpoint relations 
in increasing degree of generality and rigour: rotational symmetry, reduction of invariants and the Wigner-Eckart theorem. 
Before doing this we briefly summarise, for the reader unfamiliar with the subject, 
the helicity formalism.

 \subsection{Helicity Formalism for $d = 6$ Effective Hamiltonian}

The description of $P \to V(\to P_1 P_2) V'$ decays in the helicity formalism goes back 
to the seminal work of  Jacob and Wick \cite{JW1959} reworked in excellent lecture notes \cite{Haber1994}. 
The  extension to $P \to V  \lone \bltwo$ or $\LAin \to \LAout \lone \bltwo $,  involving  an auxiliary  timelike polarisation, has been known  for some time e.g.  
\cite{Tub02} and  \cite{Gutsche:2013pp}.  The extension beyond the dimension six Hamiltonian  has been worked out  in full generality  in \cite{GHZ2016}. 
Here we focus on the helicity formalism as applied to the dimension six effective 
Hamiltonian which contains operators of the type $H_{\textrm{eff}}^{d=6}  
\supset \bar  \qout  \ga_\al \qin \, \bar  \ell_1 \ga^\al \ell_2$. 
In what follows it is sufficient to introduce the following six operators for the hadronic part 
\begin{equation}
\label{eq:OP}
O_{S(P)}  \equiv  \bar \qout  (\ga_5) \qin  \;,\quad
O^\mu_{V(A)} \equiv     \bar \qout \ga^\mu (\ga_5) \qin  \;,\quad  
O^{\mu\nu}_{T_{(5)} } \equiv  \bar \qout   \sigma^{\mu \nu}  (\ga_5) \qin \;, 
\end{equation}
and analogous ones for the leptons.
 In the absence of QED-corrections, the matrix element factorises into 
 a hadronic and a leptonic part
 \begin{alignat}{2}
&  \matel{\LAouts \, \ell_1 \bar{\ell_2} }{H^{\qin \to \qout  \ell^+\ell^-}_{\textrm{eff}}}{\LAinb}  
   =   & &  \sum_{i = S,P}  \matel{ \LAouts }{ H_i }{\LAinb}   \matel{\ell_1 \bar{\ell_2}}{L_i}{0} \;  +  
   \nonumber \\[0.1cm]
 &   & & 
 \sum_{i = V,A}  \matel{ \LAouts }{ H_i^\mu }{\LAinb}   \matel{\ell_1 \bar{\ell_2}}{L_i^{\mu'}}{0}   g_{\mu \mu'}  \; +
  \nonumber \\[0.1cm]
 &   & & 
 \sum_{i=T,T_5} \matel{ \LAouts }{ H_i^{\mu \nu} }{\LAinb} \matel{\ell_1 \bar{\ell_2}}{L_i^{\mu'\nu'}}{0}g_{\mu \mu'}g_{\nu \nu'}  \;, 
 \end{alignat}
 where the hadronic current is $H^{\mu ..}_i \propto O^{\mu ..}_i$ in \eqref{eq:OP} 
 and the leptonic currents are given by $L_S = \bar \ell_1 \ell_2$ etc. The discussion in oversimplified in that we do not discuss different chiralities but there is no loss in doing so.
 
 The key  insight in applying  the helicity formalism to effective field theories comes from 
 the observation that the  Minkowski metric can be replaced by the
 sum of the usual helicity vectors plus an auxiliary timelike polarisation $\boost(q,t) = q/\sqrt{q^2}$, transverse to the physical polarisation vectors 
(for which  $q \cdot \boost(q,\pm,0) =0$).
The completeness relation then reads
 \begin{equation}
\label{eq:c-relation}
\sum_{\la,\la' \in \{t, \pm,0\}  }  \boost^{\mu}(q,\la) \boost^{*\nu}(q,\la')  G_{\la \la' }
= g^{\mu\nu} \;, \quad G_{\la \la' } = \text{diag}(1,-1,-1,-1) \;,
\end{equation}
where the first entry in $G_{\la \la' }$ refers to $\la = \la' = t$ and a complete set of 
helicity vectors is given in \APP \ref{app:PV}.  In this way  Minkowski indices are traded 
for helicity indices.   This then defines the helicity amplitude (HA) characterised 
by  its helicities.  Note that in the decay one helicity can be eliminated by 
generalised helicity conservation. 
In this counting the timelike helicity counts as zero since it does 
not transform under a rotation in the plane transverse to the movement \cite{HZ2013,GHZ2016}. 
We choose the spin quantisation axis to be the $q$-direction in the $\LAinb$ restframe.
Parameterising the \HA  as  
\begin{alignat}{1}
\label{eq:Hschema}
&  \HAhad{\spinq }{\nega{\spinLap} }{V}  \equiv   \matel{ \LAouts(\spinLap) }{ H^V_\mu }{\LAinb(\spinLab) } \, \boost^{*\mu}(q,\spinq) \;, 
\end{alignat}
the  
helicity conservation equation reads
\begin{equation}
\label{eq:h-con}
\spinLab  =  \spinq + \nega{\spinLap}\;  \;,
\end{equation}
where the notation $\nega{\spinLap} \equiv - \spinLap$ is employed throughout 
and  $ \spinq$ is the helicity of the fictitious particle in the decay 
chain $ \LAinb \to  \LAouts + q( \to \lone \bltwo)$.
The lepton helicity amplitudes follow in the same way
$\HAlep{ \la_1}{ \la_2  }{V}  \equiv   \matel{ \lone \bltwo }{ L^V_\mu }{0 } \, \boost^{\mu}(q,\spinq)$ but are not the focus of this paper and their explicit expression can be found in \APP A.3 in \cite{GHZ2016}.

 \subsection{Three Pathways to Endpoint Relations}
 
In explaining endpoint relations in terms of
rotational symmetry, invariants and the Wigner-Eckart theorem
it is useful to write 
an explicit parametrisation of the momenta involved 
\begin{equation}
\momLab^\mu = (m_{\LAinb},0,0,0) \;, \quad \momLapboth^{\mu} = (E_{\LAouts(\LAstarouts)},0,0,-\vv)  \;,  \quad 
 q^{\mu} = (q_0 ,0,0,\vv) \;,
\end{equation}
where $\momLab \equiv  \momLapboth + q$, $q_0 = \sqrt{q^2 +\vv^2}$, 
$E_{\LAouts(\LAstarouts)} = \sqrt{\massLapsboth^2+  \vv^2}$  and 
\begin{equation} 
\label{eq:vv}
\vv  \equiv \frac{\lambda^{1/2}(\massLab^2,\massLapsboth^2,q^2)}{2\massLab} \;,
\end{equation}
is the velocity in the $\LAinb$ restframe 
(with $\la(x,y,z) = x^2 +y^2 + z^2 - 2 x z - 2 x z - 2 y z$ the K\"all\'en-function).
 The $\vv \to 0$ limit makes the symmetry enhancement explicit: all $4$-momenta 
 are proportional to each other. The insightful   viewpoint is that it is the external momenta $\vv \neq 0$ that breaks the rotational symmetry. 
Upon discussing the rotational- and invariant-viewpoints it is convenient 
to consider $B(\momLab) \to K^*(\momLap) (\ga^* (q)\to \ell^+ \ell^-) $ 
as this avoids  dealing with Dirac spinors \cite{HZ2013}.  This is not an issue for the Wigner-Eckart viewpoint  \cite{Z13} which can be considered as one of its strengths. 
\begin{itemize}
\item \emph{Rotational Symmetry:} 
In the case of $B \to K^* \ell^+ \ell^-$, $\spinLab =0$ and 
$\spinLap = \spinq$ by \eqref{eq:h-con}. Hence the entire \HA is characterised by the helicity of the $K^*$  and one may write  $H^V_{\la_{K^*}}$.  
At the kinematic endpoint  rotational symmetry is restored and no direction 
is distinct from the viewpoint of the  $K^*$-meson. Hence the \HAs must be degenerate 
as found in the explicit evaluation at the level of the form factors (FFs)  
$H_0^i = - H^i_+ = - H^i_- $ where $i = V,A$.\footnote{\label{foot:tensor} For the tensor the same helicity structure emerges (cf. \SEC B in \cite{HZ2013}) $H_{t0}^T = - H^T_{t+} = - H^T_{t-}$ and 
$H_{+-}^T = - H^T_{+0} = + H^T_{-0}$, where we correct a sign error in 
Eq.(13) in \cite{HZ2013}, and  remind the reader that $t$ counts as zero helicity in the conservation equation.}
The respective signs
depend on the conventions of the phases to which we will return to shortly.
\item \emph{Reduction of Invariants:}  A common way to count structures is 
to write down the number of invariants that can be formed. At the kinematic endpoint
there is only a  single non-vanishing invariant, namely $\eta^*(\momLap,\spinLab) \cdot \boost^*(q,\spinq)$. The momenta cannot be involved in scalar products with the polarisation vectors 
since they are vanishing by the transversity. The Levi-Civita tensor also vanishes as 
there are only three independent vectors. Hence there is only one structure contributing and the \HAs are therefore degenerate.
\item  \emph{Wigner-Eckart theorem:} 
The most powerful method is the application of the Wigner-Eckart theorem 
due to (restored)  rotational symmetry. The helicity information is then 
solely governed by the Clebsch-Gordan coefficients (CGC) and the reduced matrix 
element contains the only dynamic information \cite{Z13}. Explicitly this statement reads 
for the case as hand (cf. \eqref{eq:Hschema})
\begin{equation}
\label{eq:master}
H^{i}_{ \spinq \nega{\spinLap}} = C^{\frac{1}{2}  1  j_{\LAouts}}_{{\spinLab} {\spinq} \nega{ \spinLap}} \, M^i_{\frac{1}{2}  1  j_{\Lambda'}} \;, \quad i = V,A,T,T_5 \;,
\end{equation}
where $C^{J j_1 j_2}_{M m_1 m_2}$ are the CGC which can be looked up in the particle 
data group resource \cite{PDG} 
and  $M$ is the reduced matrix element,  independent of the helicity quantum numbers. 
It seems worthwhile to point out that whereas  \eqref{eq:master} always holds it does,
as we shall see explicitly,  hold trivially in half of the cases ($M =0$) by parity conservation of QCD.
\end{itemize}
Equipped with \eqref{eq:master} we can then quote our main results
\begin{equation}
\label{eq:main12}
\frac{1}{2}^+ \to \frac{1}{2}^{\pm}:  \quad
H_{0 \frac12 }^i : H_{ 1 \nega{\frac12} }^i  =   
 C^{\frac{1}{2}1  \frac{1}{2} }_{\frac{1}{2} 0 \frac{1}{2}} : 
C^{\frac{1}{2} 1  \frac{1}{2} }_{{\frac{1}{2}} 1 \nega{\frac{1}{2}} }  = 1 : - \sqrt{2} \;,
\end{equation} 
and by parity the non-vanishing \HAs are $i ={A,T_5(V,T)}$ (cf. \SEC\ref{sec:BFF} just below for clarifications of the notation) and
 \begin{alignat}{1}
\label{eq:main32}
& \frac{1}{2}^+ \to \frac{3}{2}^{\pm}: 
 H_{0 \frac12 }^i : H_{ 1 \nega{\frac12} }^i : 
H_{ \nega{1} {\frac32} }^i    =  
    C^{\frac{1}{2}1 \frac{3}{2} }_{\frac{1}{2}0\frac{1}{2} }  :  C^{\frac{1}{2}1 \frac{3}{2} }_{\frac{1}{2}{1}\nega{\frac{1}{2}} } : C^{\frac{1}{2} 1 \frac{3}{2} }_{\nega{\frac{1}{2}}\nega{1} \frac{3}{2}} = 
1 :
 -   \sqrt{\frac{1}{2}}
   :  - \sqrt{\frac{3}{2}}  \;,
\end{alignat} 
$i = {A,T_5(V,T)}$ for the spin-$\frac{3}{2}$ baryon. In the remaining part of this paper we
will verify these relations at the level of the short distance form factors and discuss the implications of the relations for the differential decay rates. 
The HAs for $\frac{1}{2}^+ \to \frac{5}{2}^{\pm}$ transitions,
which are of phenomenological interest,
 vanish at the endpoint by angular momentum conservation.  They would not vanish if 
 the transition operators were of spin 2 (cf. section 5.2 in  \cite{GHZ2016}).
  
 \section{The Baryon Decays $\frac{1}{2}^+ \to \frac{1}{2}\left( \frac{3}{2}\right)^{\pm}  \ell_1 \bar{\ell_2}$
 }
 \label{sec:baryons}
 
 We choose to illustrate the endpoint relations in the baryonic decays by means 
 of explicit  form factors. We stress, once more, that the endpoint relations go beyond and also apply to any long distance matrix elements such as for example quark-loops originating 
 from $4$-quark operators of the $\bar u  c \bar q q$-type.  
 
 \subsection{The Baryonic Form Factors}
\label{sec:BFF}

As for baryon form factors there does not seem to be a standard form factor 
basis unlike in the $B \to V$ case where the basis in \cite{Wirbel:1985ji,BSZ2015} 
is used in most analysis. The baryon form factors can be divided into helicity and non-helicity-based ones. We will choose to introduce a basis which is not helicity-based as this makes the analytic structure clearer and endpoint relations are satisfied automatically 
without the need to impose further constraints cf. \SEC\ref{sec:constraints}.
We do follow the logic of the basis in  \cite{Wirbel:1985ji} in that the scalar form factor is separated and the constraint from the Ward identity or equation of motion is implemented 
by a relation to a single axial/vector form factor. 
Using our code we have checked that we agree with \cite{Gutsche:2017wag} 
in all HAs  and with the vector HAs in  \cite{Descotes-Genon:2019dbw} using their definitions.


\subsubsection{The  $\LAin_{J^P = \frac{1}{2}^+} \to \LAout_{J^P = \frac{1}{2}^{\pm}}$ Form Factors} 
\label{sec:311}

For the $\LAouts(\frac{1}{2}^+)$ final state, we define the dimensionless \FFs by
\begin{alignat}{2}
\label{eq:FF12}
&   \matel{ \LAouts(\momLap,\spinLap) }{O_{V(A)}^\mu}  {\LAinb(\momLab,\spinLab) } 
&\;=\;&
\bar u  (\ga_5)
\Big[P_1^\mu \fV_1^{V(A)} (q^2) + 
    P_2^\mu  \fV_2^{V(A)} (q^2)
    + P_0^\mu \fV_0^{V(A)}   (q^2)
    \Big] u \;, \nonumber \\[0.1cm] 
    &  \matel{ \LAouts(\momLap,\spinLap) }{O^{\mu}_{T_{(5)}}
      }  {\LAinb(\momLab,\spinLab) } 
&\;=\;&
\bar u  (\ga_5)
\Big[P_1^\mu \fT_1^{V(A)}  (q^2)+
    P_2^\mu  \fT_2^{V(A)}  (q^2)
    \Big]
    u \;,
\end{alignat}
which is an alternative to \eqref{eq:FF12}
\begin{equation}
\label{eq:P12}
P_0^\mu =  \frac{{\hat{q}}^\mu}{\hat{q}^2} \slashed{\hat{q}} \;, \quad
P_1^\mu =  \hat{q}^2 \gamma^\mu  - {{\hat{q}}^\mu} \slashed{\hat{q}}  \;, \quad 
P_2^\mu =  - i {\sigma^{\mu \hat{q}}} \;,
\end{equation}
with  $ \sigma^{\mu \hat{q}} \equiv \sigma^{\mu\nu}  \hat{q}_\nu$, 
hatted quantities are understood to be divided by $\massLab$ (i.e. $\hat{q} \equiv q/\massLab$), $\bar u \equiv \bar u(\momLap, \spinLap)$, $u \equiv u(\momLab,\spinLab)$ and we define the amendment $\LAouts(\frac{1}{2}^+) \to N(1535)(\frac{1}{2}^-)$ by 
$\bar u \to  \bar  u  \ga_5$ on the right-hand side.
The introduction of 
\begin{equation}
\label{eq:hatFF}
 \fV_1(q^2) \equiv  \frac{\fVbar_1(q^2)}{\hat{q}^2} \;,
 \end{equation}
will prove useful for implementing the regularity constraint.

Since $q \Cdot P_{1,2} =0$, $\fV_0$ is the scalar \FF.
 The non-vanishing scalar matrix elements are given by
\begin{alignat}{2}
\label{eq:FF12scalar} 
& ( m_{\qin}-m_{\qout})   \matel{ \LAouts_{J_P = \frac{1}{2}^+}(\momLap,\spinLap) }{\bar \qout \qin }  {\LAinb(\momLab,\spinLab) } 
&\;=\;& (   m_{\LAinb}- m_{\LAouts} )  \bar u  u 
 \fV_0^{V}   (q^2)  \;, \nonumber \\[0.1cm]
 & (m_{\qin}+ m_{\qout} )   \matel{ N_{J_P = \frac{1}{2}^-}(\momLap,\spinLap) }{\bar \qout \ga_5 \qin }  {\LAinb(\momLab,\spinLab) } 
&\;=\;& (m_{\LAinb} + m_{\LAouts}     )  \bar u \ga_5 u 
 \fV_0^{A}   (q^2)  \;.
 \end{alignat}
There are further constraints as for the $B \to V$ form factors.  The analogue of 
$A_0^{B \to V} (0) = A^{B \to V}_3(0)$ and $T_1^{B \to V}(0) = T_2^{B \to V}(0)$ are  
$\fV^{V(A)}_0(0) = \fVbar^{V(A)}_1(0)$ 
and $\fT_2^V(0) = \fT_2^A(0)$ respectively.  The former assures regularity and the second one is an algebraic constraint that follows from Dirac algebra cf.  \APP\ref{app:t12}.
Summarising one finds the following constraints
\begin{alignat}{2}
\label{eq:12constraints}
& \frac{1}{2}^+ \to \frac{1}{2}^{\pm}:\quad  & &    \fV^{V(A)}_0(0) = \fVbar^{V(A)}_1(0)  \;,     \quad \; \fT_2^V(0) = \fT_2^A(0) \;, 
\end{alignat}
on the \FFs which must be automatically satisfied by any computation. 
This makes up a total of ten independent \FFs with three additional constraints at $q^2 =0$.

\subsubsection{The  $\LAin_{J^P = \frac{1}{2}^+} \to \LAstarout_{J^P = \frac{3}{2}^{\pm}}$ Form Factors}

The discussion of the excited baryon \FFs proceeds very much in analogy to 
\SEC\ref{sec:311}.
The main difference is that there's an additional structure and this gives four 
additional \FFs\!\!.
For the $\LAstarouts(1535)(J^P= \frac{3}{2}^-)$ final state, we define  dimensionless \FFs   by\footnote{Contraction with $q$ does not lead to any loss of information at the level of the helicity amplitudes as there are only six of them cf. footnote \ref{foot:tensor}. 
How this works out precisely has been  elucidated recently 
 in \cite{Papucci:2021pmj}.}
\begin{alignat}{2}
\label{eq:FF32}
& \matel{\LAstarouts(\momLap,\spinLap) }{O^\mu_{V(A)}}{\LAinb(\momLab,\spinLab) }  
&\;=\;& 
\bar \RS_\al  (\ga_5)
\Big[ {\cal P}_1^{\al\mu} \FV_1^{V(A)} +{\cal P}_2^{\al\mu} \FV_2^{V(A)}+ {\cal P}_3^{\al\mu} \FV_3^{V(A)}+{\cal P}_0^{\al\mu} \FV_0^{V(A)}
\Big]
u \;, \nonumber \\[0.1cm]
& \matel{\LAstarouts(\momLap,\spinLap) }{O^\mu_{T_{(5)}}}{\LAinb(\momLab,\spinLab) }  &\;=\;& 
\bar \RS_\al (\ga_5)
\Big[ {\cal P}_1^{\al\mu} \FT_1^{V(A)} +{\cal P}_2^{\al\mu} \FT_2^{V(A)}+ {\cal P}_3^{\al\mu} \FT_3^{V(A)}
\Big]
u \;, 
\end{alignat}
where the amendment $\LAstarouts(1535)(J^P= \frac{3}{2}^-)  \to 
\LAstarouts(1720)(J^P = \frac{3}{2}^+)$ is defined by  
$\bar{\RS}  \to \bar{\RS}   \ga_5  $ on the right-hand side.
Above  $\FV_i \equiv \FV_i(q^2)$, $\FT_i \equiv \FT_i(q^2)$, $u \equiv u(\momLab,\spinLab)$ $ \RS_\al \equiv  \RS_\al(\momLapstars,\spinLap)$ is a $\frac{3}{2}$-spinor where $\al$ is a Lorentz index, 
satisfying 
$\ga^\al \RS_\al = 0$ and $\momLapstars^\al \RS_\al = 0$ cf. \APP\ref{app:32}.
The Lorentz structures ${\cal P}_i$ read
\begin{alignat}{4}
\label{eq:P32}
&  {\cal P}_0^{\al \mu} &\;=\;&  \frac{\hat{q}^\al  \hat{q}^\mu}{\hat{q}^2}  \;, \quad 
&  &  {\cal P}_1^{\al \mu} &\;=\;& 
\hat{q}^{\al}  \momLapstarsh^\mu-   ( \momLapstarsh \Cdot \hat{q})    g^{\al \mu}   \;, \quad \nonumber  \\[0.1cm]
&  {\cal P}_2^{\al \mu}  &\;=\;& \hat{q}^\al   \ga^\mu -  \slashed{\hat{q}}  g^{\al \mu}\;, \quad 
& &  {\cal P}_3^{\al \mu}  &\;=\;& \hat{q}^2 g^{\al \mu} - \hat{q}^\al  \hat{q}^\mu    \;,
 \end{alignat} 
 where $\momLapstarsh \Cdot \hat{q} = \frac{1}{2}(1 - \massLapstarsh^2-\hat{q}^2)$.
 Similar to before since $q_\mu {\cal P}^{\al \mu}_{1,2,3} =0$, $\FV_0$ is the scalar form factor. The non-vanishing scalar matrix elements are 
\begin{alignat}{2}
\label{eq:FF32scalar}
& ( \hat{m}_{\qin}- \hat{m}_{\qout} )   \matel{ \LAstarouts_{J_P = \frac{3}{2}^+}(\momLapstars,\spinLap) }
{\bar \qout \qin }  {\LAinb(\momLab,\spinLab) } 
&\;=\;& \hat{\momLab} \Cdot   \bar \RS \,  u  \,
  \FV_0^{V}   (q^2)  \;, \nonumber \\[0.1cm]
 & (\hat{m}_{\qin}+ \hat{m}_{\qout} )   \matel{ \LAstarouts_{J_P = \frac{3}{2}^-}(\momLapstars,\spinLap) }
 {\bar \qout \ga_5 \qin }  {\LAinb(\momLab,\spinLab) } 
&\;=\;& 
\hat{\momLab} \Cdot   \bar \RS  \, \ga_5 u \,
 \FV_0^{A}   (q^2)  \;.
 \end{alignat}
There are further constraints as for the $B \to V$ form factors. The regularity constraint at 
$q^2 =0 $ is $\FV_0^{A(V)}(0) = \FVbar_3^{A(V)}(0)$ with the hatted form factor analogous as in \eqref{eq:hatFF},
and the analogue of $T_1^{B \to V}(0) = T_2^{B \to V}(0)$  
is the  double constraint $\FT^V_{1}(0) = \FT^A_{1}(0)$ and  
$ \FT^A_{2}(0) = \FT^V_{2}(0) +  \massLapstarsh  \FT^V_{1}(0) $ 
which we have derived using all the seven   Lorentz structures  
(cf. \APP\ref{app:t32}) and thereby go beyond the current literature.
This makes a total of fourteen independent \FFs with four additional constraint at $q^2 =0$. In summary  the following constraints
\begin{alignat}{1}
\label{eq:32constraints}
&  \frac{1}{2}^+ \to \frac{3}{2}^{\mp}: 
 \FV_0^{V(A)}(0) = \FVbar_3^{V(A)}(0)   \, , \;
 \FT^V_{1}(0) = \FT^A_{1}(0)  \,, \;  
 \FT^{A(V)}_{2}(0) = \FT^{V(A)}_{2}(0) +  \massLapstarsh  \FT^{V(A)}_{1}(0)   \;,
\end{alignat}
apply and must hold independent of the method of computation.

\subsection{The Baryonic Helicity Amplitudes  and Verification of Endpoint Relations}

In this section we give the explicit  \HAs in terms of form factors.  
Being more explicit than previously, the \HAs read
\begin{alignat}{3}
& H^{V(A)}_{\spinq \nega{\spinLap} } &\; \equiv\;&  \matel{\LAstaroutboth(\momLapboth,\spinLap) }{O^\mu_{V(A)}}{\LAinb(\momLab,\spinLab) } & &  \boost^{*}_\mu(q,\spinq) \;, \nonumber \\[0.1cm]
& H^{T(T_5)}_{\spinq \nega{\spinLap} } &\; \equiv\;&  \matel{\LAstaroutboth(\momLapboth,\spinLap) }{O^\mu_{T_{(5)}}}{\LAinb(\momLab,\spinLab) }  & & \boost^{*}_\mu(q,\spinq) \;.
\end{alignat}
The following kinematic abbreviations are helpful
\begin{alignat}{2}
\label{eq:abbrev}
& \Mplmi \equiv \massLab \pm \massLapsboth \;, \quad 
& & \Q{\pm} \equiv \Mplmi^2 - q^2 \;, \quad  \nonumber \\[0.1cm]
& \lambda(\massLab^2,\massLapsboth^2,q^2)  =    \Q{+} \Q{-} = {4\massLab^2 \vv} \;, 
\quad & & \DelM = \massLab^2  - \massLapsboth^2 \;,
\end{alignat}
where $\la$ is  the K\"all\'en-function, defined above \eqref{eq:vv}.
 Moreover, the substitutions at the kinematic endpoint
are as follows 
\begin{equation}
 q^2 \to \Mmi^2 \;, \quad 
 \sqrt{\Q{+}} \to 2 \sqrt{\massLab \massLapsboth } \;, \quad 
\sqrt{\Q{-}} \to  \sqrt{\frac{\massLab}{\massLapsboth}} \vv \;.  
\end{equation}

\subsubsection{The $\LAin_{J^P = \frac{1}{2}^+} \to \LAout_{J^P = \frac{1}{2}^{\pm}}$  Helicity Amplitudes}
\label{sec:HA12}

Using the explicit \FF parameterisation in \eqref{eq:FF12}, for $ \frac12^+ \to \frac{1}{2}^{+}$, we get the following expressions
\begin{alignat}{3}
\label{eq:end12}
& \hat{H}_{ 0\frac 1 2 }^{V(A)} &\;=\;& \sqrt{\hat{q}^2 \Qh{\mp} } \,
\Big(  \Mplmih  f_1^{V(A)} +     f_2^{V(A)}  \Big)  &\;\stackrel{\textrm{endpoint}}{\rightarrow} \;& 
\left\{\begin{array}{l} 
 \phantom{- \sqrt{2}}  \vec{f} \cdot  \vec{v}_{0\frac{1}{2}}\,  c_{1/2} \hat{\vv}
   \\[0.3cm]
\phantom{- \sqrt{2} } \hat{\EPR}^A_{1/2}     \end{array} \right.   
\;,  \nonumber \\[0.1cm]
& \hat{H}_{ 1 \nega{\frac{1}{2}}}^{V(A)} &\;=\;&  \sqrt{2 \Qh{\mp}} \,
\Big(  \hat{q}^2  f_1^{V(A)}  + \Mplmih  f_2^{V(A)} \Big) &\;\;\stackrel{\phantom{\textrm{endpoint}}}{\rightarrow}  \;& 
\left\{\begin{array}{l}   - \sqrt{2}  \vec{f} \cdot  \vec{v}_{ 1 \nega{\frac{1}{2}}} \,  c_{1/2} \hat{\vv} \\[0.3cm]
- \sqrt{2}  \hat{\EPR}^A_{1/2}     \end{array} \right.  \;, 
\end{alignat}for 
the  vector and the axial  HAs,  evaluated to leading order in $\vv$ on the right-hand side at the kinematic endpoint with
\begin{equation}
 \hat{\EPR}^A_{1/2}  \equiv  { 2 \Mmih \sqrt{  \massLapsh }}{} ( 
  \Mmih f_1^A(\Mmi^2)  +    f_2^A(\Mmi^2) ) \;,
\end{equation}
the auxiliary velocity vectors read ($ c_{1/2} \equiv 1/\sqrt{{\massLapsh}}$)
\begin{equation}
\vec{v}_{0\frac{1}{2}} = (  \Mplh \Mmih , \;\Mmih     ) \;, \quad   
\vec{v}_{ 1 \nega{\frac{1}{2}}}   =  - (   \Mmih^2 , \;\Mplh     ) \;,
\end{equation} 
and the scalar product is to be interpreted as a two dimensional one in Euclidean space 
with $\vec{f} = (f_1,f_2)$.
For the timelike \HA we obtain 
\begin{alignat}{3}
\label{eq:end12time}
& \hat{H}_{ t\frac12}^{V(A)} &\;=\;&  \pm  \sqrt{\frac{\Qh{\pm}}{\hat{q}^2}} \,
 \Mmiplh \, \fV_0^{V(A)}  &\;\to \;& 
\left\{\begin{array}{l}  2 \sqrt{\massLapsh}\,  \fV_0^V(\Mmi)   \\[0.2cm]
- \frac{1}{\sqrt{\massLapsh}}  { \frac{\Mplh}{\Mmih}} \,  \fV_0^A(\Mmi)  \, \hat{\vv} 
\end{array} \right.   \;.
\end{alignat}

All other \HAs follow by substitution rules from  the ones above.
Firstly,   the tensor amplitudes follow from the vector amplitudes by  replacing  $(\fV_0,\fV_1,\fV_2) \to (0,\fT_1,\fT_2)$. 
Second,
\begin{equation} 
J^P = \frac{1}{2}^+: \quad \HAhad{\nega{\spinq} }{\nega{\spinLap}}{V(A)} = \pm 
\HAhad{{\spinq} }{\spinLap}{V(A)} \;,  \qquad 
J^P = \frac{1}{2}^-: \quad  \HAhad{\nega{\spinq} }{\nega{\spinLap}}{V(A)} = \mp 
\HAhad{{\spinq} }{\spinLap}{V(A)} \;.
\end{equation}
Third, the  $ \frac{1}{2}^+ \to \frac{1}{2}^{-}$ \HA are obtained by replacing  
$ \fV^{V(A)}_i ,  \fT^{V(A)}_i $ by $ \fV^{A(V)}_{i},  \fT^{A(V)}_i $ which follows from the
definition \eqref{eq:FF12}.
At last we observe that   the endpoint relations \eqref{eq:main12},
\begin{equation}
\label{eq;end12}
 \hat{H}_{ 0 {\frac{1}{2}}}^{A(T_5)}
 = - \frac{1}{\sqrt{2}} \hat{H}_{ 1 \nega{\frac{1}{2}}}^{A(T_5)}  \;,
\end{equation}
 is obeyed as dictated by the Wigner-Eckart theorem.

\subsubsection{The $\LAin_{J^P = \frac{1}{2}^+} \to \LAstarout_{J^P = \frac{3}{2}^{\mp}}$  Helicity Amplitudes}
\label{sec:HA32}

Employing the  \FF parameterisation in \eqref{eq:FF32}, for $ \frac12^+ \to \frac{3}{2}^{-}$, the following expressions are obtained 
\begin{alignat}{3}
\label{eq:end32}
& \hat{H}_{0\frac12}^{V(A)} &\;=\;& 
\pm \frac{\sqrt{2   \hat{q}^2 \Qh{\pm} } }{\sqrt{3}} \left(  (\pm \massLapstarsh ) 
\FV^{V(A)}_1 \pl  
 \FV^{V(A)}_2   
 \pl  \frac{q^2  \mi \DelMh }{2 \sqrt{2}(\pm  \massLapstarsh)  }  \FV_3^{V(A)}    \right)   &\;\to \;& 
\left\{\begin{array}{l} \phantom{- \sqrt{\frac{1}{2}}}  \hat{\EPR}^V_{3/2}
   \\[0.3cm]
  \phantom{- \sqrt{\frac{1}{2}}}  \vec{F} \cdot  \vec{v}_{0\frac{1}{2}}  \,  c_{3/2}  \hat{\vv}     \end{array} \right.   
\;,  \nonumber \\[0.1cm]
& \hat{H}_{ 1 \nega{\frac{1}{2}}}^{V(A)} &\;=\;& 
 \frac{\sqrt{ \Qh{\pm} } }{\sqrt{3}} 
\Big(  \frac{q^2 \mi \DelMh}{2}    \FV_1^{V(A)}  +   \frac{q^2 \mi \Mmiplh  }{(\pm  \massLapstarsh)  }    \FV_2^{V(A)}  \pl  \hat{q}^2 \FV_3^{V(A)} \Big) &\;\to \;& 
\left\{\begin{array}{l}  - \sqrt{\frac{1}{2}} \hat{\EPR}^V_{3/2}
  \\[0.3cm]
  - \sqrt{\frac{1}{2}} \vec{F} \cdot  \vec{v}_{1\nega{\frac{1}{2}}} \, c_{3/2} \hat{\vv}     \end{array} \right. 
 \;,  \nonumber \\[0.1cm]
 & \hat{H}_{ \nega{1} {\frac{3}{2}}}^{V(A)} &\;=\;& \sqrt{\Qh{+}}  \left( \frac{\hat{q}^2 \mi \DelMh}{2}
  \FV_1^{V(A)} \mi   \Mmiplh   
 \FV_2^{V(A)} \pl  \hat{q}^2  
 \FV_3^{V(A)}  )\right)   &\;\to \;& 
\left\{\begin{array}{l}    - \sqrt{\frac{3}{2}}   \hat{\EPR}^V_{3/2} 
  \\[0.3cm]
 - \sqrt{\frac{3}{2}} \vec{F} \cdot  \vec{v}_{\nega{1}{\frac{3}{2}}} \, c_{3/2} \hat{\vv}   \end{array} \right. \;, 
\end{alignat}
where the endpoint expression reads 
\begin{equation}
 \hat{\EPR}^V_{3/2}  \equiv  \sqrt{\frac{8 \massLapstarsh}{3}} {\Mmih }  ( \massLapstarsh  \FV_1^V( \Mmi^2)  \pl \FV_2^V( \Mmi^2) \mi\Mmih \FV_3^V( \Mmi^2) ) \;,
\end{equation}
the auxiliary velocity vectors are ($c_{3/2} \equiv \frac{\sqrt{2}\Mmih}{\sqrt{3 \massLapstarsh} }$)
\begin{equation}
 \vec{v}_{0{\frac{1}{2}}} =  ( \massLapstarsh , \mi 1,    \mi \Mmih ) \;, \quad
  \vec{v}_{1\nega{{\frac{1}{2}}} } =  ( \massLapstarsh ,\mi  \frac{3 \pl \massLapstarsh}{1\mi \massLapstarsh},   \mi \Mmih ) \;, \quad
 \vec{v}_{\nega{1}{\frac{3}{2}}} =  ( \massLapstarsh , \frac{1 \pl \massLapstarsh}{1\mi \massLapstarsh},   \mi \Mmih ) \;,  
\end{equation}
and the scalar product is to be interpreted as for the $\frac{1}{2}$-case. 
For the timelike \HA we obtain 
\begin{alignat}{3}
\label{eq:end32time}
& \hat{H}_{ t \frac12}^{V(A)} &\;=\;& - 
\sqrt { \frac{{\Qh{
\mp} }   }{ {6 \hat{q}^2} }} \frac{\Qh{\pm}}{\massLapstarsh  }  \, \FV_0^{V(A)}  &\;\to \;& 
\left\{\begin{array}{l}  - 2 \sqrt{\frac{2}{3 }}    \frac{1}{\massLapstarsh^{1/2} \Mmih} 
\FV_0^V(\Mmi^2)
\, \hat{\vv}   \\[0.2cm]
 -  \sqrt{\frac{2}{3 }}    \frac{1}{\massLapstarsh^{3/2} \Mmih} \FV_0^A(\Mmi^2) \, \hat{\vv}^2 
   \end{array} \right.   \;,
\end{alignat}
The substitution rules for the other  \HAs are as follows.
Firstly,  the tensor amplitudes follow from the vector amplitudes by  replacing
$(\FV_0,\FV_1,\FV_2,\FV_3 ) \to (0,\FT_1,\FT_2,\FT_3)$. Second, 
\begin{equation}
J^P = \frac{3}{2}^-: \quad \HAhad{\nega{\spinq} }{\nega{\spinLap}}{V(A)} = \pm 
\HAhad{{\spinq} }{\spinLap}{V(A)} \;,  \qquad 
J^P = \frac{3}{2}^+: \quad  \HAhad{\nega{\spinq} }{\nega{\spinLap}}{V(A)} = \mp 
\HAhad{{\spinq} }{\spinLap}{V(A)} \;.
\end{equation}
Third, the  $ \frac{1}{2}^+ \to \frac{3}{2}^{+}$ \HA are obtained by the 
replacement
 $  \FV^{V(A)}_{i}, \FT^{V(A)}_{i}  \to   \FV^{A(V)}_{i}, \FT^{A(V)}_{i}$.
 The endpoint relations \eqref{eq:main32},
\begin{equation}
 \hat{H}_{ 0 {\frac{1}{2}}}^{V(T)} = 
- \sqrt{2}  \hat{H}_{ 1 \nega{\frac{1}{2}}}^{V(T)} =   
 - \sqrt{\frac{2}{3}}  \hat{H}_{ \nega{1} {\frac{3}{2}}}^{V(T)}  \;,
\end{equation}
 are satisfied by the explicit expression in \eqref{eq:end32}.

\subsection{Specific Endpoint Features compared with $B \to V \lone \bltwo$}

In this section we compare the endpoint behaviour
of the $\frac{1}{2}$- and $\frac{3}{2}$-\HAs, of the previous section, with the ones for 
$B \to V \lone \bltwo$. Generally, parity determines 
whether the power of the velocity $\vv$ is even or odd.  This is the appeal of the so-called
\begin{equation}
H_{\perp(\para)} = \frac{1}{\sqrt{2}} ( H_{1} \mp H_{\nega{1}}) \;, \quad H_{0}   = H_0 \;,
\end{equation}
transversity basis  (the subscript on the right-hand side corresponds to the helicity index $\spinq$).
Now the $H_{\perp,\para,t}$ \HAs, but not $H_0$, are of definite parity and this leads to 
selection rules in the power of the velocity  collected in \TAB\ref{tab:kappa} 
for the two types baryonic modes  compared  against $0^- \to 1^{\pm}$ which includes 
$B \to V \lone \bltwo$ discussed in our previous work.
 
An interesting feature is that in the $\frac{1}{2} \to \frac{1}{2}$ mode the timelike \HA 
\eqref{eq:end12time}
does not vanish at the endpoint unlike in the other cases. The reason for this is that the mode can be in an $S$-wave ($l=0$), w.r.t  the lepton pair, whereas the other ones require 
a $P$-wave ($l =1$).  For most of phenomenology this term 
however remains unimportant since it is suppressed by  lepton masses. 

In the context of $B \to V \lone \bltwo$  we advocated a $\vv$-expansion which was based 
on the universality of terms linear in $\vv$. 
Inspecting the results in \EQs\eqref{eq:end12}
and \eqref{eq:end32} one observes that this universality no longer holds for baryons. 
The reason for it is that there  are several \FFs for each parity. 
This is in contradistinction to  $B \to V \lone \bltwo$ where 
the vector operator $O_V^\mu$ \eqref{eq:OP} 
gives rise to  a single \FF, denoted by $V^{B \to V}(q^2)$.
 Hence in the baryonic case the form factor dependence does not drop when ratios of HAs, linear in $\vv$, are taken.  From this viewpoint the $\vv$-expansion in $B \to V \lone \bltwo$ might be seen as somewhat accidental.

\begin{table}[t]
  \centering
  \begin{tabular}{l   |  l    |    l     |  ll   |  l      }
  $J^P \to {J'}^{P'}$          &    $H_\perp$ &    $H_\para$    &
  $H_0$ &   & $H_t$       \\[0.1cm] \hline   
  $\frac{1}{2}^+  \to     \frac{1}{2}^{\pm} $         &  $H^{A(V)} \propto\vv^{0(1)}$  & 
  $H^{V(A)}\propto\vv^{1(0)}$ &  $H^{V}\propto \vv^{1(0)}$ & $H^{A}\propto \vv^{0(1)}  $    & $H^{V(A)}\propto \vv^{0(0)}$    \\[0.1cm] 
   $\frac{1}{2}^+  \to     \frac{3}{2}^{\mp} $         &  $H^{A(V)}\propto\vv^{1(0)}$  & 
  $H^{V(A)}\propto\vv^{0(1)}$ & $H^{V}\propto \vv^{0(1)}$ & $ H^{A}\propto \vv^{1(0)}  $ 
  &  $H^{V(A)}\propto \vv^{1(1)} $    \\[0.1cm] 
 $0^-  \to     1^{\pm} $         &  $H^{A(V)}\propto\vv^{0(1)}$  & 
  $H^{V(A)}\propto\vv^{1(0)}$   &  $H^{V(A)}\propto \vv^{1(0)} $ & & 
  $H^{V(A)} \propto \vv^{1(1)} $  
  \end{tabular}
  \caption{\small  Leading $\vv$-dependance in $J^P \to {J'}^{P'}$ HAs. 
  The table is an extension of \TAB 1 in \cite{HZ2013} where 
  $C_\pm$ corresponds to $H^{V(A)}$ for $0^-  \to     1^{-} $ which was the main focus 
  of that paper. Changing the parity of the initial state demands $V \leftrightarrow A$ in the table above.  \HAs which are not given, such as 
  $H^A_{\perp}|_{\frac{1}{2}^+ \to \frac{1}{2}^-}$, do vanish by parity in QCD. 
  The  \HA $H_0$ is not of definite parity and this is why both vector and axial ones contributes to the same power. The $T(T_5)$ \HAs follows the same pattern as the ones 
  for $V(A)$ respectively.}
  \label{tab:kappa}
\end{table}

\section{Constraints on Helicity Based Form Factors}
\label{sec:constraints}

Before applying our results   to particular 
angular distributions in the literature we are inclined to clarify a point which 
does not seem to have received specific  attention 
in the literature.\footnote{The discussion in  \cite{Descotes-Genon:2019dbw}  
has not been worked out into concrete constraints. Additionally, in our opinion, 
the  dependence on $\Q{\pm}$ from the spinors should not enter the discussion 
as they are external objects to the correlation function.} 
This concerns constraints on helicity-based form factors which are
often  used when discussing specific angular distributions e.g.  \cite{BDF2015} and \cite{Descotes-Genon:2019dbw}. 

The helicity-based form factors have been introduced  in \cite{Feldmann:2011xf} 
for the $\frac{1}{2}^+ \to \frac{1}{2}^+$ case and 
extended to $\frac{1}{2} \to \frac{3}{2}$ in references \cite{Meinel:2020owd,Descotes-Genon:2019dbw}.
The basic idea is that each HA is proportional to a single form factor. 
This of course raises the question of how the  endpoint relations 
\eqref{eq:main12} and \eqref{eq:main32} can be satisfied.
This can only be remedied if  there are  constraints between 
the helicity form factors. 
This is indeed the case as the price for introducing 
helicity-based form factors is that the analytic structure of the form factors and the \HAs 
differ.  
This is due to  $\frac{1}{s_\pm}$-poles in the analogue of the Lorentz structures \eqref{eq:P12} and \eqref{eq:P32}.  Now, since by first principles 
there cannot be any poles of the form $\frac{1}{\Q{\pm}}$ in the HA, we are to conclude that  there must be constraints.\footnote{A pole would correspond to a particle of mass $ \Mplmi^2$ 
by  the  dispersion relation or LSZ formalism e.g. 
chapter 10  in \cite{Weinberg:1995mt}.   Note though,  that for  
there can be pole singularities at  $q^2 = \Mpl^2$ in  the form factor.  
There can only 
but  branch cuts,   corresponding to the threshold of the 
$\ell^+ \ell^- \to \LAinb \LAstaroutboth$ process related by crossing symmetry. 
At $q^2 =  \Mmi^2$ there are no singularities of this type on the first sheet but there are branch cuts on the second sheet. 
These type of singularities are known as pseudo-thresholds \cite{Eden:1966dnq}. 
For a simple example thereof we refer the reader to  the lectures notes \cite{Zwicky:2016lka}. 
In summary whereas  there are singularities at $\Mplmi$, in terms of branch cuts,   pole singularities are absent because there is no known (stable) particle in QCD of this mass and quantum numbers.} 

For completeness and clarity we reproduce the helicity-based form factor definition \cite{Feldmann:2011xf}, 
an alternative to \eqref{eq:FF12},  but using the notation in \cite{BDF2015} in \SEC 3\footnote{The definition is  more 
straightforward in \cite{Feldmann:2011xf} but a  change of notation is  was 
undertaken from \cite{Feldmann:2011xf} to \cite{BDF2015}: 
$f(g)_+ = f_0^{V(A)}$ and $f(g)_\perp = f_\perp^{V(A)}$.} 
The constraint for the $\frac{1}{2} \to \frac{1}{2}$-case has been derived in \cite{Detmold:2015aaa} indirectly by using 
a  similar basis to ours and requiring them to match with the  helicity form factor based results.
\begin{alignat}{2}
\label{eq:FF32}
&   \matel{ \LAouts(\momLap,\spinLap) }{O_{V(A)}^\mu}  {\LAinb(\momLab,\spinLab) } 
&\;=\;&
\pm \bar u  (\ga_5)
\Big[ 
    Q_0^\mu  f_0^{V(A)} (q^2)
    + Q_\perp^\mu f_\perp^{V(A)}   (q^2) + Q_t^\mu f_t^{V(A)} (q^2) 
    \Big] u \;, \nonumber \\[0.1cm] 
    &  \matel{ \LAouts(\momLap,\spinLap) }{O^{\mu}_{T_{(5)}}
      }  {\LAinb(\momLab,\spinLab) } 
&\;=\;&
\pm \bar u  (\ga_5)
 \Big[Q_0^\mu f_0^{T(T_5)}  (q^2)+
    Q_\perp^\mu  f_\perp^{T(T_5)}  (q^2)
    \Big]
    u \;,
\end{alignat}
where the Lorentz structures are
\begin{equation}
\label{eq:Q}
Q_t^\mu = \Delta_{\mp} \frac{q_\mu}{q^2} \;, \quad 
Q_0^\mu = \frac{\Delta_\pm}{s_\pm}(p \pl  p_p \mi  \frac{\Delta^2}{q^2} q)^\mu \;,
\quad Q_\perp^\mu = (\ga \mp \frac{2 m_p}{s_\pm} p_p  - \frac{2 m_{\Lambda_c}}{s_\pm}p)^\mu \;, 
\end{equation}
with abbreviation defined in \eqref{eq:abbrev} and with some mild abuse of notation 
the first and second entry correspond to $V,T$ and $A,T_5$ respectively.
Now, by requiring the absence of poles at $q^2 = \Mmi^2 \equiv \qmaxsq$ one find the following constraints
\begin{equation}
\label{eq:con-12}
\frac{1}{2}^+ \to \frac{1}{2}^{\pm}:  f_0^{A(V)}( \qmaxsq) =  f_\perp^{A(V)}( \qmaxsq)  \;.
\end{equation}
All results in this section are valid for the tensor form factors as well with 
the usual replacements $V,A \to T,T_5$.   
The further constraint at $q^2 = \Mpl^2$ reads 
\begin{equation}
\label{eq:con-12b}
\frac{1}{2}^+ \to \frac{1}{2}^{\pm}:  f_0^{V(A)}( \Mpl^2) =  f_\perp^{V(A)}( \Mpl^2) \;,
\end{equation}
and can be obtained by interchanging $V \leftrightarrow A$ and $m_p \leftrightarrow - m_p$ 
from the one at $q^2 = \Mmi^2$.
Applying \eqref{eq:con-12} to the expressions in \EQ 3.13 in \cite{BDF2015} one does 
indeed find that $H^A_0(+1/2,+1/2) :  H_+^A(-1/2,+1/2) =  1 : -\sqrt{2}$ in agreement 
with our generic prediction  \eqref{eq:main12}.\footnote{We do not seek to find agreement in signs since this depends on the conventions of the authors.  
Of course these sign cancel  when combined with  the specific angular decay rates, provided everything was handled consistently. Nevertheless or 
somewhat accidentally  signs agree.}

Again, for completeness and clarity we reproduce the $\frac{1}{2} \to \frac{3}{2}$ 
helicity-based form  factors, using notation in   \cite{Descotes-Genon:2019dbw} except introducing the 
dimensionless form factor 
$\bar f_g = f_g/ \massLab$, 
\begin{alignat}{2}
\label{eq:FF32}
& \matel{\LAstarouts(\momLap,\spinLap) }{O^\mu_{V(A)}}{\LAinb(\momLab,\spinLab) }  
&\;=\;& 
\bar \RS_\al  (\ga_5)
\Big[ p^\al(Q_0^{\mu} f_0^{V(A)} +Q_\perp^{\mu} f_\perp^{V(A)} +Q_t^{\mu} f_t^{V(A)})
+ Q_g^{\al\mu} \bar{f}_g^{V(A)}
\Big]
u \;, \nonumber \\[0.1cm]
& \matel{\LAstarouts(\momLap,\spinLap) }{O^\mu_{T_{(5)}}}{\LAinb(\momLab,\spinLab) }  &\;=\;& 
\bar \RS_\al (\ga_5)
\Big[ p^\al(Q_0^{\mu} f_0^{T(T_5)} +Q_\perp^{\mu} f_\perp^{T(T_5)} 
+ Q_g^{\al\mu} \bar{f}_g^{T(T_5)}
\Big]
u \;, 
\end{alignat}
with $Q^\mu$ as defined in \eqref{eq:Q} and 
\begin{equation}
Q_g^{\al\mu}  = m_{\Lambda_c} \big( g^{\al \mu} - \frac{m_N p^\al}{s_\mp}( \ga \mp \frac{2 p_p}{m_N} \pm 2 \frac{m_N p \pm  m_{\Lambda_c} p_p}{s_\pm})^\mu \big) \;, 
\end{equation}
with the same meaning for the first and second entry as previously.
Proceeding along the same reasoning as before, one gets 
the following constraints at $q^2 = \Mmi^2 \equiv \qmaxsq$
\begin{alignat}{2}
\label{eq:con-32}
& \frac{1}{2}^+ \to \frac{3}{2}^{\mp}: \quad  & &   f_\perp^{V(A)}( \qmaxsq) = - 
 \frac{  \massLab \massLapstars \bar{f}_g^{V(A)}(  \qmaxsq) }{\Q{-}} \;,  \nonumber \\[0.1cm]
&  &  & f_0^{V(A)}( \qmaxsq) =    \frac{2 \massLab  \massLapstars  \bar{f}^{V(A)}_g(  \qmaxsq) }{\Q{-}} \frac{ \massLab -  \massLapstars }{ \massLab +  \massLapstars} \;,  \nonumber \\[0.1cm]
& & & f_\perp^{A(V)}( \qmaxsq) = f_0^{A(V)}( \qmaxsq) - \frac{\bar{f}_g^{A(V)}( \qmaxsq)}{4} \;.
\end{alignat}
Since $\bar{f}_g(  \qmaxsq) $ is generically  non-vanishing this means that the 
 $f_{0,\perp}$ form factors are singular at the endpoint and can be seen as a
 justification for introducing the slightly different notation as in \cite{Meinel:2020owd}.
 The constraint at $q^2 =  \Mpl^2$ reads 
 \begin{alignat}{2}
\label{eq:con-32b}
& \frac{1}{2}^+ \to \frac{3}{2}^{\mp}: \quad  & &   f_\perp^{A(V)}( \Mpl^2) =  
 \frac{  \massLab \massLapstars \bar{f}_g^{A(V)}(  \Mpl^2) }{\Q{+}} \;,  \nonumber \\[0.1cm]
&  &  & f_0^{A(V)}( \Mpl^2) =  -   \frac{2 \massLab  \massLapstars  \bar{f}_g^{A(V)}(  \Mpl^2) }{\Q{+}} \frac{ \massLab +  \massLapstars }{ \massLab -  \massLapstars} \;,  \nonumber \\[0.1cm]
& & & f_\perp^{V(A)}( \Mpl^2) = f_0^{V(A)}( \Mpl^2) - \frac{\bar{f}_g^{V(A)}( \Mpl^2)}{4} \;,
\end{alignat}
and can again be obtained by interchanging $V \leftrightarrow A$ and $\massLapstars \leftrightarrow 
- \massLapstars$.
Applying \eqref{eq:con-32} to \EQ 3.8 in  \cite{Descotes-Genon:2019dbw} we do find that 
$H_0^V(+1/2,+1/2) : H^V_+(1/2,-1/2) : H^V_+(1/2,3/2)   =  1 :  -   \sqrt{\frac{1}{2}} :  - \sqrt{\frac{3}{2}} $ in accordance with \eqref{eq:main32}.
In summary, as expected, once the constraints are applied the endpoint relations are satisfied.

\subsection{Comparison with the Literature}

From the viewpoint of theory endpoint relations of HAs are valuable 
for consistency checks of  computations and valid fit parameterisations in connections  with form factor constraints.  The concrete use for experiment is discussed in the
upcoming \SEC\ref{sec:vic}.

Since we have clarified how endpoint relations work it seems worthwhile 
 to mention a few papers that could benefit from applying this knowledge for checks 
 and parameterisations. 
 For example,  the results in  \cite{Boer:2018vpx}  
 for both $\frac{1}{2}^+ \to \frac{1}{2}^+$ and 
 $\frac{1}{2}^+ \to \frac{3}{2}^-$  do not satisfy the endpoint relations (even when the subleading Isgur-Wise function is set to zero). There are minor issues in 
 \cite{Bordone:2021bop} fo $\Lambda_b \to \Sigma$.
 The lattice form factor fit for $ \Xi_{{c}} \to \Xi$  \cite{Zhang:2021oja}  
 needs the constraints \eqref{eq:con-12}  imposed in order to satisfy the endpoint relations.
We do agree with the 
$\frac{1}{2}^+ \to \frac{3}{2}^-$ results in  \cite{Descotes-Genon:2019dbw} 
 and the endpoint relation are satisfied once the constraints \eqref{eq:con-32} are imposed.

\section{Angular Distributions}
\label{sec:AD}

We provide a brief analysis of angular distributions in the limit of zero lepton mass 
based on the references \cite{BDF2015} and \cite{Descotes-Genon:2019dbw} 
for the $\frac{1}{2}^+ \to \frac{1}{2}^{+} $ and $\frac{1}{2}^+ \to \frac{3}{2}^{-} $ cases.\footnote{The adaption to the other parity final states is relatively straightforward when 
taking into account the previously stated transformation rules.}
We define the normalised distributions as follows
\begin{alignat}{2}
& \KN(q^2, \theta_\ell, \theta_\Lambda, \phi)  &\;=\;& \left(\frac{d \Gamma}{dq^2}  \right)^{-1}
 \frac{d^4 \Gamma(\Lambda_b \to \Lambda( \to N \pi) \ell^+ \ell^-)}{d q^2\,d \cos\theta_\ell\,d \cos\theta_\Lambda\,d \phi}  \;, \nonumber \\[0.1cm]
& \LN(q^2, \theta_\ell, \theta_\Lst, \phi)  &\;=\;& \left(\frac{d \Gamma}{dq^2}  \right)^{-1}
\frac{d^4\Gamma(\Lambda_b \to \Lambda(1520)( \to N \bar K) \ell^+ \ell^-)}{dq^2d\cos{\theta_\ell}d\cos{\theta_\Lst}d\phi} \;,
\end{alignat}
where we refer to the references above for the definition of the decay angles.
In what follows we assume that the initial baryon is unpolarised. 
A salient difference between the two decays is that the $\Lambda$ is stable in QCD 
and decays via the  weak decay $\Lambda \to N \pi$, parameterised by the parity-violating parameter $\al$.
In what follows we introduce the notation $\vev{ f(\theta_\ell, \theta^{(*)}_\Lambda, \phi)}$ 
\begin{equation}
\vev{ f(\theta_\ell, \theta^{(*)}_\Lambda, \phi) } = \int_{-1}^{1} d \cos \theta_\ell 
\int_{-1}^{1} d \cos \theta^{(*)}_\Lambda \int_0^{2\pi} d \phi  \,f(\theta_\ell, \theta^{(*)}_\Lambda, \phi)  \;,
\end{equation}
to stand for the averaging over the angles such that 
\begin{alignat}{2}
& \vev{\KN} &\;=\;&   \frac{\normi}{3} ( \KN_{1cc} + 2 \KN_{1ss})   = 1 \;,   \nonumber \\[0.1cm]
& \vev{\LN} &\;=\;&  \frac{\normi}{9} ( \LN_{1cc} + 2 (\LN_{1ss} + \LN_{2cc} + 2 \LN_{2ss} + \LN_{3ss})) =  1 \;,
\end{alignat}
 are normalised to unity 
and $\Omega \equiv \vev{1} = 8 \pi $ is the phase space normalisation factor.

\subsection{The $\LAin_{J^P = \frac{1}{2}^+} \to \LAout_{J^P = \frac{1}{2}^{\pm}}$  Angular Distribution at the Endpoint}
\label{sec:AD12}

For the $\frac{1}{2}^+ \to \frac{1}{2}^{+}$ case we reproduce for convenience 
the distribution found in \cite{BDF2015}
\begin{equation}
\begin{aligned}
    \label{eq:angular-distribution}
     \KN(q^2, \theta_\ell, \theta_\Lambda, \phi)    & =
         \big(  \KN_{1ss} \sin^2\theta_\ell +\,  \KN_{1cc} \cos^2\theta_\ell +  \KN_{1c} \cos\theta_\ell\big) \,\cr
    &  + \big(  \KN_{2ss} \sin^2\theta_\ell +\,  \KN_{2cc} \cos^2\theta_\ell +  \KN_{2c} \cos\theta_\ell\big) \cos\theta_\Lambda
    \cr
    &  + \big(  \KN_{3sc}\sin\theta_\ell \cos\theta_\ell +  \KN_{3s} \sin\theta_\ell\big) \sin\theta_\Lambda \sin\phi\cr
    &  + \big(  \KN_{4sc}\sin\theta_\ell \cos\theta_\ell +  \KN_{4s} \sin\theta_\ell\big) \sin\theta_\Lambda \cos\phi  \;,
\end{aligned}
\end{equation}
for which $\Lambda_b \to \Lambda (\to N \pi) \ell^+ \ell^-$  was considered in specifically.  
The distribution including initial state polarisation can be found in \cite{Blake:2017une}.
All terms involving $\theta_\Lambda$ are then proportional to the parity-violating 
parameter 
\begin{equation}
\al =  \frac{| H_{0\nega{\frac12}}|^2- | H_{0{\frac12}}|^2 }{| H_{0\nega{\frac12}}|^2 + | H_{0{\frac12}}|^2 } \;,
\end{equation}
and some values are known: e.g. $\al_{\Lambda \to p \pi^-} = 0.732 (14)$ \cite{PDG}.  Note that if 
the final state baryon decays via the strong force then $\al \to 0$.
Using our results, 
the particular  
expressions found in \SEC\ref{sec:constraints}, and inserting them into their expressions  for the $K$'s we find that at the endpoint 
\begin{alignat}{2}
\label{eq:Kend12}
& \{ \KN_{1ss} ,  \KN_{1cc}    \} 
&\; \to\;& \norma \{ 1,1\}\;,   \nonumber \\[0.1cm]
&  \{   \KN_{2c} ,  \KN_{4s} \} 
&\; \to\;& \frac{ \al r} {\normi }  \{ - 1, 1 \}\;,  \quad r  \approx 
 \frac{ 2 \Rea[  (C_{9} - C'_{9})(C_{10} - C'_{10})^*]}{ | C_{9} - C'_{9}|^2+|C_{10} - C'_{10}|^2} \;,
\end{alignat}
where other terms are vanishing because either the amplitudes in question vanish 
or they vanish as  all  HAs are proportional to each other.  
The latter point implies that
the $K's$ proportional to imaginary parts of products of amplitudes vanish.
The factor $r$, which receives additional  correction from dipole operators which are however negligible in the 
Standard Model, is a curiosity of the parity violation in the decay which would not be present if the $\Lambda$ would decay via the strong force ($\al=0$). 
Parity conservation of QCD assures that endpoint relations are more effective in 
constraining the angular distribution at the endpoint.
Moreover, if the decay was adapted to $\frac{1}{2}^+   \to \frac{1}{2}^-$ then 
${C' \to - C'}$ which makes it manifest that this is a parity violating channel.
Finally the angular 
distribution in \eqref{eq:angular-distribution} assumes the simple form 
\begin{equation}
 \KN(q^2, \theta_\ell, \theta_\Lambda, \phi)   \to 
\norma  \left( 1  -  \al r (\cos \theta_\ell \cos \theta_\Lambda  - \sin \theta_\ell \sin \theta_\Lambda \cos\phi) \right) \;,
\end{equation}
towards the kinematic endpoint of zero $\Lambda$-velocity in the $\Lambda_b$ 
restframe.
As a simple application we may compute the longitudinal lepton polarisation 
\begin{equation}
\label{eq:FL}
 F_L\big |_{\frac{1}{2}^+ \to \frac{1}{2}^{+}}  \equiv \vev{ \omega_L(\theta_\ell) } =  \frac{ |H_{0 \frac{1}{2} }|^2  }{ |H_{0 \frac{1}{2} }|^2 + |H_{ 1 \nega{\frac{1}{2} }}|^2}   = \frac{\normi}{3} (2 \KN_{1ss} - \KN_{1cc}  )
   \to \frac{1}{3} \;,
\end{equation}
where $\omega_L(\theta_\ell) = 2 - 5 \cos^2 \theta_\ell$. 
In  \cite{BDF2015,Descotes-Genon:2019dbw} this observable is denoted $F_0$ and $F_1 = 1 -F_0 \to \frac{2}{3}$ and   $F_H = \frac{3}{2} F_1 = \frac{3}{2}(1-F_L) \to 1$ 
in the notation of \cite{Boer:2018vpx}.  Other examples are the forward backward asymmetries
\begin{alignat}{1}
& A^{\ell}_{FB} \big |_{\frac{1}{2}^+  \to \frac{1}{2}^+ } \equiv \vev{ \textrm{sgn} \cos \theta_\ell}  \;=\;  
\frac{\normi}{2}\KN_{1c}  \;\to\;  0 \;,  \nonumber \\[0.1cm]
& A^{\Lambda}_{FB} \big |_{\frac{1}{2}^+  \to  \frac{1}{2}^{+}}\equiv \vev{ \textrm{sgn} \cos\theta_\Lambda }  \;=\;  
\frac{\normi}{6}(\KN_{2cc}+ 2 \KN_{2ss}) \;\to\;  0 \;,  \nonumber \\[0.1cm]
& A^{\Lambda \ell}_{FB} \big |_{\frac{1}{2}^+  \to  \frac{1}{2}^{+}}\equiv \vev{ \textrm{sgn}(  \cos\theta_\ell\cos\theta_\Lambda ) } \;=\;  
\frac{\normi}{4}\KN_{2c} \;\to\;  - \frac{1}{4} \al r \;.
\label{eq:12-endpoint}
\end{alignat}
As in $B \to V\ell^+ \ell^+$ the forward backward asymmetries vanish at the endpoint except the combined one which is sensitive to the parity violating decay of the 
$\Lambda$-baryon.
 For the $\frac{1}{2}^+ \to \frac{1}{2}^{+}$ case into dineutrinos, as in $\Lambda_b \to \Lambda (\to N \pi) \nu \bar \nu$, the   angular analysis is limited to
\begin{align}
\left(\frac{d \Gamma}{dq^2}  \right)^{-1}
 \frac{d \Gamma}{d \cos\theta_\Lambda} = 
  \hat{K}_{1cc} + 2 \hat{K}_{1 ss} +  (\hat{K}_{2 cc}+ 2 \hat{K}_{2 ss})\cos\theta_\Lambda \to 3 \;,
\end{align}
  which assumes a constant value at the endpoint.
Generally one can extract $A^{\Lambda}_{FB}$, which vanishes at the endpoint  as predicted in (\ref{eq:12-endpoint}).

\subsection{The $\LAin_{J^P = \frac{1}{2}^+} \to \LAout_{J^P = \frac{3}{2}^{\pm}}$  Angular Distribution at the Endpoint}
\label{sec:AD32}

For the $\frac{1}{2}^+ \to \frac{3}{2}^{-}$ decay 
the distribution found in  \cite{Descotes-Genon:2019dbw}  is
\begin{equation}\label{eq:angobs}
\begin{aligned}
 \LN(q^2, \theta_\ell, \theta_\Lst, \phi)
 & = \cos ^2\theta_\Lst \left(\LN_{1c} \cos \theta_\ell+\LN_{1cc} \cos ^2\theta_\ell+\LN_{1ss} \sin ^2\theta_\ell\right)\\
&+ \sin ^2\theta_\Lst \left(\LN_{2c} \cos
   \theta_\ell+\LN_{2cc} \cos ^2\theta_\ell+\LN_{2ss} \sin ^2\theta_\ell\right)\\
&+ \sin ^2\theta_\Lst \left(\LN_{3ss} \sin ^2\theta_\ell \cos^2
   \phi+\LN_{4ss} \sin ^2\theta_\ell \sin \phi \cos
   \phi\right)\\
   &+\sin \theta_\Lst \cos \theta_\Lst \cos \phi \left(\LN_{5s} \sin \theta_\ell+\LN_{5sc} \sin \theta_\ell \cos \theta_\ell \right)\\
   &+\sin \theta_\Lst \cos \theta_\Lst\sin \phi \left(\LN_{6s} \sin
   \theta_\ell+\LN_{6sc} \sin \theta_\ell \cos \theta_\ell \right) \;,
\end{aligned}
\end{equation}
for which $\Lambda_b \to \Lambda (\to N \bar K) \ell^+ \ell^-$  was specifically considered. 
Proceeding in the same way as before using our findings, in \SEC\ref{sec:constraints}, 
and inserting them into their expressions for the $L$'s we get the following
\begin{equation}
\label{eq:Lend32}
\{ \LN_{1cc}, \LN_{1ss}, \LN_{2cc}, \LN_{2ss}, \LN_{3ss}, \LN_{5sc} \} \to 
\norma \left\{\frac{1}{2},\frac{5}{4},\frac{5}{4},\frac{5}{4},
  - \frac{3}{4},-\frac{3}{2}\right\} \;,
\end{equation}
non vanishing contributions which are consistent with the partial list  given in  \EQ4.14
 \cite{Descotes-Genon:2019dbw}.
 Again the one proportional to the imaginary part vanish 
since all amplitudes are proportional to each other.
With \eqref{eq:Lend32} the angular distribution assumes the simple form
\begin{alignat}{2}
& \LN \to
\norma (\frac{1}{4}(5 - 3    \sin^2 \theta ^*_{\Lambda }   \cos^2 \phi)
\mi   \frac{3}{8} \sin 2  \theta_{\ell}   \sin 2 \theta ^*_{\Lambda } \cos \phi    \mi  
\frac{3 \cos \theta^2_\ell}{4} (\cos^2 \theta ^*_{\Lambda } \mi   \sin^2 \theta ^*_{\Lambda } \cos^2 \phi  ))  \;,
\end{alignat}
at the kinematic endpoint of zero $\Lambda^*$-velocity  in the restframe of
the decaying particle. As a simple illustration let us consider the longitudinal polarisation 
\begin{alignat}{2}
& F_L \big |_{\frac{1}{2}^+ \to \frac{3}{2}^{-}} \equiv \vev{ \omega_L(\theta_\ell) } &\;=\;&   \frac{ |H_{0 \frac{1}{2} }|^2  }{ |H_{0 \frac{1}{2} }|^2 + |H_{ 1 \nega{\frac{1}{2} }}|^2   + |H_{ \nega{1} {\frac{3}{2} }}|^2}    \nonumber \\[0.1cm]
& &\;=\;& \frac{\normi}{9} ( 2  \LN_{1ss} - 2 \LN_{2cc} + 4 \LN_{2ss} + 2\LN_{3ss} - \LN_{1cc})
   \to \frac{1}{3} \;,
\end{alignat}
for which the same numerical result is obtained as in the $\frac{1}{2}$-case and thus 
$F_T = 1- F_L  \to \frac{2}{3}$ (and $F_H = \frac{3}{2} F_T \to 1$). Again the forward backward asymmetry in the lepton angle
 \begin{equation}
A_{FB} \big |_{\frac{1}{2}^+ \to \frac{3}{2}^{-}} \equiv \vev{ \textrm{sgn} \cos \theta_\ell  }  =  \frac{\normi}{6}( \LN_{1c}  + 2 \LN_{2c}) \to  0 \;,
\end{equation}
vanishes. The forward backward asymmetries involving the Baryon angle (cf. \eqref{eq:12-endpoint}) 
vanish,  $A^{\Lambda}_{FB} \big |_{\frac{1}{2}^+  \to  \frac{3}{2}^{-}} =0$ and 
$A^{\Lambda \ell}_{FB} \big |_{\frac{1}{2}^+  \to  \frac{3}{2}^{-}} =0$, since the final state Baryon decays via the strong force.
The  angular analysis for the $\frac{1}{2}^+ \to \frac{3}{2}^{-}$ case into dineutrinos as in $\Lambda_b \to \Lambda (\to N \bar K) \nu \bar \nu$ 
 is restricted to
\begin{align}
\left(\frac{d \Gamma}{dq^2}  \right)^{-1} \frac{d \Gamma}{d  \cos \theta_\Lst } =  (  \hat{L}_{1 cc}+ 2 \hat{L}_{1 ss}) \cos^2  \theta_\Lst
+( \hat{L}_{2 cc} + 2 \hat{L}_{2 ss} + \hat{L}_{3 ss}) \sin^2  \theta_\Lst  \rightarrow 3 \, ,
\end{align}
and assumes a constant value at the endpoint.

\subsection{In the Vicinity of the Endpoint}
\label{sec:vic}

Endpoint relations are exact at the kinematic endpoint. However, 
since experiment provides bins in the  $q^2$-variable, the question of how 
 endpoint prediction deviate in the vicinity of the endpoint poses itself.
We propose to take inspiration from the inclusive $b \to s \ell^+ \ell^-$ decay.
Without QED-correction, the inclusive differential  rate, $d \Gamma / (d q^2 d \cos \theta_\ell)$, consists of three terms: the total rate, $A_{\textrm{FB}}$ and $F_L$  \cite{Fukae:1998qy,Lee:2006gs} out of which $F_L$ is the most interesting for our purpose. 
Since we aim at a rough estimate we restrict ourselves to the tree-level prediction 
for which
 \begin{equation}
F_L(s,\hat{m})=\frac{s-1 - \hat{m} ^4+\hat{m} ^2 (s+2)}{(s-1)(2s+1) - \hat{m} ^4- \hat{m} ^2 (s-2)} 
\;, 
\end{equation}
 $s_{\text{max}}=(1-\hat{m})^2$,  $\hat{m} = m_b/m_s$ for  $b \to s$  
 and accordingly for other quark level transitions.
 Note that $F_L(s_{\text{max}}) = 1/3$ which is just a special case of the $\frac{1}{2} \to \frac{1}{2}$-transition \eqref{eq:FL}. Defining  
 $R_L(s,\hat{m}) = (F_L(s,\hat{m})-F(s_{\text{max}}))/F(s_{\text{max}})$,  one gets
 the relative deviations 
\begin{alignat}{2}
\label{eq:RL}
& (R_L(0.8 s_{\text{max}} ,\frac{m_{d,u}}{m_{b,c}} ),R_L(0.9 s_{\text{max}} ,\frac{m_{d,u}}{m_{b,c}})) &\;\approx \;&(0.15,0.07) \;, \quad  \nonumber \\[0.1cm]
& (R_L(0.8 s_{\text{max}} ,\frac{{m}_s}{m_b}),R_L(0.9 s_{\text{max}} ,\frac{{m}_s}{m_b})) &\;\approx \;&(0.18,0.09) \;, \quad  \nonumber \\[0.1cm]
& (R_L(0.8 s_{\text{max}} ,\frac{{m}_s}{m_c}),R_L(0.9 s_{\text{max}} ,\frac{{m}_s}{m_c})) &\;\approx \;&(0.22,0.11) \;, \quad  \nonumber \\[0.1cm]
& (R_L(0.8 s_{\text{max}}  ,\frac{{m}_c}{m_b}),R_L(0.9 s_{\text{max}} ,\frac{{m}_c}{m_b}) &\;\approx \;&  (0.27,0.13) \;,
\end{alignat}
with $m_{d,u}/m_{b,c} \approx 0$, 
$ {m}_s/m_b \approx  0.02 $, $m_s/m_c \approx 0.08$ and $m_c/m_b \approx 0.28$ for the $20\%$- and $10\%$-bin towards the endpoint. It is noted that mass effects are sizeable.
In what follows we distinguish the semileptonic- and the FCNC-case as the latter are  
plagued by broad resonances towards the endpoint. 
\begin{itemize}
\item \emph{Semileptonic-case:} For $b \to u,c \to d$ and $c \to s$
we may take the first and third line \eqref{eq:RL} at face  value and conclude that for a bin at the endpoint, with  $(20, 10)\%$ of the entire range,  the deviation from the endpoint value to be  $(8,4)\%$  and $(11,6)\%$ respectively. 
For $b \to c$ we see that the effect is  roughly doubled in size, as compared to the $u,d$-transitions, with $(16,8)\%$ for the  $(20,10)\%$-bin  respectively.
\item \emph{FCNC-case:}  the $b \to d,c \to u$  and $b \to s$ transitions include 
in the vicinity of the endpoint
broad resonances with potentially different phases. Based on the plot in Fig.11 in 
\cite{LZ2014} (see also  \cite{Brass:2016efg}) we estimate for beauty that the
deviation will be $50\%$ on top of the semileptonic-case. 
This is  $(12,6)\%$ and $(14,7)\%$  for the $(20,10)\%$-range intervals.\footnote{The situation in charm changes drastically  in the presence  of new physics, which is much less constrained than in beauty. 
In fact $F_L$  tests the Standard Model \cite{Golz:2021imq}.
A concrete BSM study for $\Lambda_c \to p \mu^+ \mu^-$ reveals a possible relative deviation in the 
$20\%$- and $10\%$-bin  of  roughly  $45 \%$ and $25 \%$, respectively.}
\end{itemize}
These estimates are understood to be valid for baryonic  as well as mesonic transitions 
and therefore fill a gap in our previous work \cite{HZ2013}
where we did not comment on this  aspect. 
 In the semileptonic case of small $\hat{m}$ the corrections are rather small 
 and
it is feasible that QED-corrections are competitive.  The subject of the latter is though 
subtle from many viewpoints. For example, 
the definition of the angular observables will depend on 
how the photon is treated within  the kinematic variables \cite{Isidori:2020acz} 
and experiment aims to remove the QED at last in the large logarithms.

\section{Summary and Conclusions}
\label{sec:conclusions}

We established endpoint relations for baryons based on 
Lorentz symmetry   
resulting  in exact ratios of helicity amplitudes at the kinematic endpoint 
cf. \EQs \eqref{eq:main12} and \eqref{eq:main32},  for the $\frac{1}{2} \to \frac{1}{2}$
and $\frac{1}{2} \to \frac{3}{2}$ baryon transitions respectively.
 These relations have been verified explicitly, in the form factor basis 
\eqref{eq:FF12} and \eqref{eq:FF32}, with resuling helicity amplitudes in \eqref{eq:end12} and \eqref{eq:end32} respectively. 
We stress that these relations are valid for long-distance matrix elements 
such as   four-quark operators (leading to resonance contributions). 
These relations have hitherto  been overlooked in the helicity-based form factor analyses
 since the   constraints established in \SEC\ref{sec:constraints} have 
 not been validated. 
 Exact angular distributions at the kinematic endpoint are derived 
for the $\frac{1}{2} \to \frac{1}{2}$
and $\frac{1}{2} \to \frac{3}{2}$ decays in \EQs\eqref{eq:Kend12} and \eqref{eq:Lend32} 
respectively. These predictions hold model-independently in the $d=6$ weak Hamiltonian. However we wish to emphasise that this framework does not cover 
light new physics (beyond the Standard Model) or QED-corrections. 

We hope that our work contributes to exciting experimental and theoretical effort in 
 baryonic modes which include FCNC and semileptonic modes such as
\begin{alignat}{1}
& \Lambda_b \to \Lambda \ell \ell \, , \quad
\Xi_b \to \Xi\ell \ell \, , \quad 
\Lambda_b \to n \ell \ell \, , \quad
\Lambda_c \to p \ell \ell \, ,  \nonumber \\[0.1cm]
& \Xi_c \to (\Sigma, \Lambda) \ell \ell \, , 
\quad 
\Omega_c \to \Xi \ell \ell \, , 
\quad
\Sigma^+ \to p \ell \ell \, ,  \nonumber \\[0.1cm]
& \Lambda_b \to \Lambda_c \ell \nu \, , \quad \Lambda_c \to \Lambda \ell \nu \, , \quad \Lambda \to p \ell \nu \, , \quad
\Xi^- \to (\Lambda, \Sigma^0) \ell \nu \;,
\end{alignat}
for $\frac{1}{2} \to \frac{1}{2}$   and\footnote{We denote with a '$\ast$' excited hadrons with spin $3/2$ and $SU(3)_F$-quantum numbers as the corresponding ground state. 
A famous example of a $\Lambda^*$ is the $\Lambda(1520)$.}
\begin{align}
\Lambda_b \to \Lambda^* \ell \ell \;,
\quad
\Omega_b \to  \Omega \ell \ell \, , \quad
\Lambda_c \to N^+ \ell \ell \;,
\quad
\Xi_c \to (\Sigma^*,\Lambda^*) \ell \ell \;,
\quad
\Omega_c \to \Xi^* \ell \ell  \;, 
\quad
\end{align}
 for the  $\frac{1}{2} \to\frac{3}{2}$ case.

\subsection*{Acknowledgements}

We are grateful to Stefan Meinel for asking us the question to what degree  endpoint relations hold for $\frac{3}{2}$-baryons, the coordination of his work 
\cite{Meinel:2021mdj}
 and for many discussions on the literature.
GH is grateful to  Marcel Golz and Tom Magorsch for discussions and coordination of their work.
RZ is supported by an STFC Consolidated Grant, ST/P0000630/1. 
The work by GH is supported in part by the Bundesministerium f\"ur Bildung und Forschung (BMBF) under project number
 05H21PECL2.

\appendix

\section{Conventions}
\label{app:conventions}

\subsection{Polarisation Vectors}
\label{app:PV}
 
We follow our conventions \cite{Z13,HZ2013,GHZ2016} for which a complete set of 
helicity vectors in the restframe is given by\footnote{
The only difference to  
\cite{Z13,HZ2013,GHZ2016} is the sign of $\gpol(\pm)$ which we adapt here to match 
the spinor conventions in \APP\ref{app:Dirac}. The correctness of the conventions can 
be established by computing 
 $$C^{\frac{1}{2} 1 \frac{1}{2}}_{\spinLab \spinq \nega{\spinLap}} \propto \bar u(\spinLap) \ga_{\mu} \ga_5 u(\spinLab) \gpol^{*\mu}(\spinq)$$ in the restframe of the spinors (for example). An alternative to the sign change  in $\gpol(\pm)$ is to introduce a  relative minus sign between the two $\la = \pm \frac{1}{2}$-spinors.} 
\begin{eqnarray}
\label{eq:JW}
\gpol(\pm)  =   -(0, \pm 1, i,0)/\sqrt{2} \;, \quad \gpol(0) = (0,0,0,1) \quad \text{and} \quad  \gpol(t) = (1,0,0,0 )  \;.
\end{eqnarray}
where  indices are understood to by raised.
 The boosted versions with respect to $q  =  (q_0,0,0, \vv) $ and  $\momLap = (E_{\LAstaroutboth},0,0,-\vv)$  ($  q_0 = \sqrt{q^2 + \vv^2}, E_{\LAstaroutboth}= \sqrt{\massLapstars^2 + \vv^2}$)
 are
 \begin{alignat}{6}
&  \boost(q,\pm) &\;=\;& \ \gpol(\pm)\;, \quad & &  \boost(q,0) &\;=\;& (\vv,0,0,q_0)\frac{1}{\sqrt{q^2}}  \;, \quad & & \boost(q,t) &\;=\;& \frac{q}{\sqrt{q^2}} \;, \nonumber \\[0.1cm]
& \aboost(\momLapstars,\pm) &\;=\;& \gpol(\mp)\;, \quad & & \aboost(\momLapstars,0) &\;=\;& (-\vv,0,0,E_{\LAstaroutboth})\frac{1}{\massLapstars} \;, \quad  & & \aboost(\momLapstars,t) &\;=\;& \frac{\momLap}{\massLapstars} \;,
\end{alignat}
All three sets of helicity vectors satisfy the completeness relation in \eqref{eq:c-relation}.
 The $\aboost$ helicity vector is needed for the $\frac{3}{2}$-spinor as shown further below.

\subsection{Dirac Algebra}
\label{app:Dirac}

We choose to work in the Dirac representation of the Clifford algebra
\begin{equation}
\label{eq:Dirac}
 \ga_0 =  \begin{pmatrix} 1&0 \\ 0& -1 \end{pmatrix}  \,, \qquad
  \ga_i = \begin{pmatrix} 0 & -\sigma^i\\ \sigma^i & 0 \end{pmatrix}\,, \qquad
  \ga_{5} = 
\begin{pmatrix} 0 \; & 1\\ 1 \;& 0 \end{pmatrix}  \;,
\end{equation}
where  $\sigma^i$ are the usual $2\times 2$ Pauli matrices $\ga_5 = i \ga^0 \ga^1 \ga^2 \ga^3$ holds, 
 $\sigma_{\mu\nu} = \frac{i}{2} [ \ga_\mu , \ga_\nu]$ and $g_{\mu\nu} = 
\textrm{diag}(1,-1,-1,-1) = \frac{1}{2} \{ \ga_\mu , \ga_\nu \}$.
With $ \varepsilon_{0123} =1$
\begin{equation}
\label{eq:5sigma}
\sigma^{\al\be} \ga_5 = -   \frac{i}{2}  \varepsilon^{\al\be\ga\de}\sigma_{\ga\de} \;,
\end{equation}
holds. The $\frac{1}{2}$-spinors (Dirac) $u$ and $v$, moving into the $z$-direction, 
are given by  (e.g. \cite{Haber1994}) 
\begin{equation}
\label{eq:uv}
u(\la) = \begin{pmatrix}  \be^+  \chi_\la \\   2 \la \be^-    \chi_\la \end{pmatrix}  \;, \quad  
v(\la)  = \begin{pmatrix}  \be^-  \chi_\la \\   - 2 \la \be^+    \chi_\la \end{pmatrix} = C \bar u^T \;,
\end{equation}
with $\chi_{\frac{1}{2}} = (1,0)$, 
$\chi_{-\frac{1}{2}} = (0,1)$ and 
$\be^{\pm} \equiv \sqrt{E \pm m}$  (with $ \Qh{\pm} =  2 (\hat{\be}^{\pm}) ^2$ in the notation of this paper). If the direction of movement is reversed then $\be^- \to -\be^-$in the formula above. For generic directions see the reference above.
In \eqref{eq:uv}  $C = i \ga^0 \ga^2$ is the charge conjugation matrix in the Dirac representation. 

\subsection{The $\frac{3}{2}$-spinor} 
\label{app:32}

The spin $3/2$-states were first discussed by Rarita and Schwinger \cite{Rarita:1941mf}.
It can be constructed out of the the spin one polarisation tensor and the 
Dirac spinor using the Clebsch-Gordan coeffcients
\begin{equation}
\label{eq:build_RS}
\RS_\al(\momLapstars,\lambda) = \sum_{\sigma = -1/2}^{1/2} C^{\frac32 \frac12 1}_{\lambda s (\lambda-\sigma)} \aboost_\al(\momLapstars,\lambda-\sigma)  u(\momLapstars,\sigma)  \;,
\end{equation}
and $\momLapstars^\al \RS_\al = 0$ holds trivially and  $\ga^\al \RS_\al = 0$ can be verified using 
the conventions used in the previous sections.  The normalisation is such that 
$\bar{\RS}^\al(\momLapstars,\lambda) \RS_\al(\momLapstars,\lambda') = - 2 \massLapstars \de_{\la,\la'}$.
Again the CGC are the ones compatible with the Condon-Shortley phase convention and can be looked  up in the \cite{PDG}.
 
 \section{The Tensor Form Factor Relation due to $\sigma^{\al\be} \ga_5 = -   \frac{i}{2}  \varepsilon^{\al\be\ga\de}\sigma_{\ga\de}$ }
 \label{app:t}
 
 In the form factor definitions \eqref{eq:FF12}  and \eqref{eq:FF32} we have contracted 
 the $\sigma_{\mu \nu}$ with $q^\nu$, as is commonly done in the literature.   
 This  leads to relations at $q^2 =0$ since the form factor without contraction is free from poles in $q^2$. In some sense these relations, which we derive below, can be seen as an artefact of the parametrisation.  
 Further note that at $q^2 =0$ the zero-helicity part decouples as one would expect from a massless particle.   
The constraints are thus necessarily between form factors in the perpendicular or transversal  directions. 
   
 \subsection{$\LAin_{J^P = \frac{1}{2}^+} \to \LAout_{J^P = \frac{1}{2}^{\pm}}$-case}
 \label{app:t12}

 Here we aim to demonstrate $\fT^V_2(0) = \fT^A_2(0)  $ of the form factors defined in 
 \eqref{eq:FF12}. The most general form of the uncontracted version of the
 tensor form factors reads
 \begin{eqnarray}
 & &  \matel{ \LAouts(\momLap,\spinLap) }
 {    \bar s i  { \sigma^{\mu \nu} } (\ga_5) b   }{\LAinb(\momLab,\spinLab) }  =  
  \\[0.1cm]  
& & \qquad    \bar u( \momLap, \spinLap) \left(    x_A(q^2)  \momLabh_{[\mu }  \hat{q}_{\nu]}
   - x_B(q^2) i \sigma_{\mu\nu} \, + 
   x_C(q^2) \ga_{[\mu }  \momLabh_{\nu ]}  + 
   x_D(q^2) \ga_{[\mu } \hat{q}_{\nu]} 
   \right)u(\momLab,\spinLab) \;, \nonumber
 \end{eqnarray}
where the square brackets denote antisymmetrisation as in $a_{[\al} b_{\be]} = 
a_\al b_\be - a_\be b_\al$. Note that due to the relation 
\begin{equation}
\label{eq:useful}
\{ \sigma_{\al \be} ,\sigma_{\mu\nu} \} = 2 \big( 
(g_{\al \mu} g_{\be \nu} - \{ \al \leftrightarrow \be\}) - i \eps_{\al\be\mu\nu} \ga_5 \big)\;,
\end{equation}
which can be derived from the Chisholm identity for example, the  $\eps_{\mu\nu p q} \bar u( \momLap, \spinLap)  \ga_5 
u(\momLab,\spinLab)$-structure can be expressed in terms of the others.
Using the identity \eqref{eq:5sigma} and the definitions 
\eqref{eq:FF12} ones arrives at the following identifications 
\begin{alignat}{4}
& \fT_1^V(q^2) &\;=\;& x_D(q^2) + \frac{1}{2} ( x_C(q^2) + \Mplh x_A(q^2) )  \;, \quad & & \fT_1^A(q^2) &\;=\;& \frac{1}{2} x_C(q^2)  \;,  \nonumber  \\[0.1cm]
& \fT_2^V(q^2) &\;=\;& x_B(q^2) + \frac{1}{2} ( \Mmih x_C(q^2) - 2 \hat{q}^2  x_A(q^2) )  \;, \quad & & \fT_2^A(q^2) &\;=\;& x_B(q^2) +  \frac{1}{2} \Mmih x_C(q^2)  \;.
\end{alignat}
Since all $x_i$, and in particular $x_A$ are non-singular for $q^2 \to 0$ it follows that
\begin{equation}
\fT_2^V(0) = \fT_2^A(0) \;,
\end{equation}
which we aimed to show.  This relation has been obtained previously in 
\cite{Gutsche:2013pp} albeit in the specific context of their computation.

 \subsection{$\LAin_{J^P = \frac{1}{2}^+} \to \LAout_{J^P = \frac{3}{2}^{\pm}}$-case}
 \label{app:t32} 

Proceeding in complete analogy we write the most generic ansatz for \eqref{eq:FF32}
with uncontracted $q$ 
 \begin{eqnarray}
   \matel{\LAstarouts(p',\spinLap) }
 {    \bar s i {\sigma^{\mu \nu} }  b   }{\LAinb(\momLab,\spinLab) }  & =& 
  \bar \RS_\al(p',\sigma) i  \left\{ \right. \!\!    \momLabh_{\al }  (
 X_A (-i  \sigma_{\mu\nu}) \pl
 X_B  \ga_{[\mu }  \momLabh_{\nu ]}  \pl 
 X_C   \ga_{[\mu } \hat{p}' _{\nu]}  \pl X_D  \momLabh_{[\mu }  \hat{p}'_{\nu]}) \nonumber \\[0.1cm]
 & +& X_E   g_{\al[ \mu} \ga_{\nu]} + X_F   g_{\al[ \mu} p_{\nu]} + X_G  g_{\al[ \mu} p'_{\nu]}  \left. \!\! \! \right \}u(\momLab,\spinLab) \;, 
 \end{eqnarray}
where here $p' = p_N$ and $X_i = X_i(q^2)$ 
for brevity and we have chosen the 
$\frac{1}{2}^- \to \frac{3}{2}^-$-transition for concreteness. 
  In reducing the structures we made use of 
\eqref{eq:useful} and the equation of motion.
Contracting with $q$ and equating with our ansatz  \eqref{eq:FF32} ones finds
\begin{alignat}{2}
& T_1^V &\;=\;&  2 X_A + \hat{\Delta}_-(X_B+X_C) + q^2 X_D + X_F +X_G \;,  \nonumber \\[0.1cm]
& T_2^V &\;=\;&  - \hat{\Delta}_+ X_A  + \frac{1}{2}(\hat{m}_N^2 - \hat{q}^2-1) X_B + 
  \frac{1}{2}(\hat{m}_N^2 + \hat{q}^2-1)   X_C  + X_E  \;,  \nonumber \\[0.1cm]
& T_3^V &\;=\;&  - X_A + \hat{\Delta}_-X_B -  \frac{1}{2}(\hat{m}_N^2 + \hat{q}^2-1)X_D- X_F
 \nonumber \\[0.1cm]
 & T_1^A &\;=\;&  2 X_A + \hat{\Delta}_ - (X_B+X_C)  + X_F +X_G \;,  \nonumber \\[0.1cm]
 & T_2^A &\;=\;&  - \hat{\Delta}_- X_A  + \frac{1}{2}( \hat{q}^2+ 2 \hat{m}_N - \hat{m}_N^2-1) (X_B+X_C)   + X_E  + \hat{m}_N (X_F+X_G) \;,  \nonumber \\[0.1cm]
& T_3^A &\;=\;&  - X_A +  \hat{m}_N(X_B + X_C)  \;.
\end{alignat}
At $q^2 =0$ the following constraints emerge
\begin{equation}
\label{eq:32T}
 \frac{1}{2}^+ \to \frac{3}{2}^{\mp}: \quad 
\FT^V_{1}(0) = \FT^A_{1}(0)  \,, \quad 
 \FT^{A(V)}_{2}(0) = \FT^{V(A)}_{2}(0) +  \massLapstarsh  \FT^{V(A)}_{1}(0)   \;,
\end{equation}
 in accordance with \eqref{eq:32constraints} and  \cite{Hiller:2007ur,Descotes-Genon:2019dbw}. The new aspect  is that the relation \eqref{eq:32T} has been derived 
 with seven  structures as  advocated in  \cite{Papucci:2021pmj}. In  
 \cite{Descotes-Genon:2019dbw} only six structures have been considered
 and nn  \cite{Hiller:2007ur}  no detail has been given.

\bibliographystyle{utphys}

\bibliography{../../Refs-dropbox/References_All.bib}

\end{document}